\documentclass[aps,preprint,groupedaddress]{revtex4}  

\usepackage{amssymb}
\pdfoutput=1
\pdfoutput=1
\pdfoutput=1
\pdfoutput=1
\pdfoutput=1
\usepackage{amsmath}
\usepackage{amssymb}
\usepackage{graphicx}
\usepackage{subfigure}
\usepackage{color}
\usepackage{mathrsfs}
\usepackage[dvipsnames]{xcolor}

\usepackage[breaklinks=true,colorlinks=true]{hyperref}
\hypersetup{colorlinks=true,citecolor=blue,linkcolor=blue,urlcolor=blue}

\usepackage[utf8]{inputenc}

\begin{document}
\immediate\write16{<<WARNING: LINEDRAW macros work with emTeX-dvivers
                    and other drivers supporting emTeX \special's
                    (dviscr, dvihplj, dvidot, dvips, dviwin, etc.) >>}

\title{Geometrically constrained sine-Gordon field: BPS solitons and their collisions}

\author{E. da Hora$^{1}$}


\author{L. Pereira$^{2}$}


\author{C. dos Santos$^{3}$}


\author{F. C. Simas$^{2}$}


\affiliation{$^{1}$Coordenação do Curso de Bacharelado Interdisciplinar em Ciência e Tecnologia, Universidade Federal do Maranhão, 65080-805, São Luís, Maranhão, Brazil.\\
$^{2}$Departamento de Física, Universidade Federal do Maranhão, Campus Universitário do Bacanga, 65085-580, São Luís, Maranhão, Brazil.\\
$^{3}$Centro de F\'isica e Departamento de F\'isica e Astronomia,
Faculdade de Ci\^encias da Universidade do Porto,
4169-007, Porto, Portugal.
}

 
\begin{abstract}


We consider an enlarged $(1+1)$-dimensional model with two real scalar fields, $\phi$ and $\chi$ whose scalar potential $V(\phi,\chi)$ has a standard $\chi^4$ sector and a sine-Gordon one for $\phi$. These fields are coupled through a generalizing function $f(\chi)$ that appears in
the scalar potential and controls the nontrivial dynamics of $\phi$. We minimize the effective energy via the implementation of the BPS technique \cite{prasom,bogo}. We then obtain the Bogomol'nyi bound for the energy and the first-order equations whose solutions saturate that bound. We solve these equations for a nontrivial $f(\chi)$. As the result, BPS kinks with internal structures emerge. They exhibit a two-kink profile, i.e. an effect due to geometrical constrictions. We consider the linear stability of these new configurations. In this sense, we study the existence of internal modes that play an important role during the scattering process. We then investigate the kink-antikink collisions, and present the numerical results for the most interesting cases. We also comment about their most relevant features.

\end{abstract}


\maketitle

\section{Introduction}
\label{intro}

Configurations with nontrivial topology appear in many areas of Physics. They are widely studied not only due to their very complicated physical properties, but also their possible applications. Commonly, they emerge in high-energy physics as a result of phase transitions. In general, they are connected to highly nonlinear models. In this case, the non-linearity is usually introduced via the potential that describes the scalar-matter self-interaction \cite{masu}.

They are also present in condensed matter physics \cite{poga,chdo}. In such a scenario, they can be studied through experimental methods. In Ref. \cite{jab}, for instance, the authors used experimental techniques to explore geometrically constrained walls that develop a well-defined two-kink profile.

The simplest topologically nontrivial structures are kinks and antikinks. They emerge in $(1+1)$-dimensional models with real scalar fields. These solutions have the same energy but topological charges with opposite signs. Under special circumstances, kinks can also be obtained via first-order differential equations that appear from the minimization of the energy via the BPS method \cite{prasom,bogo}. Naturally, the complexity of the theory depends on the number of the scalar sectors.

The scattering of kinks has been object of special interest. Regarding this, several studies have identified the occurrence of resonance windows and chaotic patterns in kink-antikink collisions \cite{sugi,camp1,anninos0}. These windows are formed as a consequence of the energy exchange between the internal modes during the scattering process. These modes are usually determined via the linear perturbation theory. By this way, kink-antikink scattering have been studied in various models, such as the non-integrable
$\phi^4$ and $\phi^6$ ones - see Refs. \cite{cape,dohamerosh,asmoraebsa} and \cite{dorosh,weig,mogasadmja}, but also in integrable scenarios like the sine-Gordon theory
\cite{blms,masd,pmrg,bgmsa}. Additional discussions can also be found in \cite{takyi,henoli,weigel,gani1,dio1,simas1,dio2,dio3,halava.2012,Lima.JHEP.2019,ivan}.

Adding more than one scalar field, the resulting theory and respective interactions become significantly more complex. For example, in Ref. \cite{alonso}, the authors explored such interactions in a model characterized by a single vacuum. The complexity has increased when two \cite{alonso2,alonso3}, three \cite{algomato} and four \cite{Halavanau,simas2022} vacuum states were included. Situations where one of the fields is in its quantum vacuum configuration were also studied, see Refs. \cite{maevvaza,mavacha}.

In the recent years, there has also been considerable research on the so-called {\textit{spectral walls}}. They are barriers formed as a result of the transition from discrete to continuous modes during kink-antikink collisions \cite{adam.2019,camoquweza}. It was established in Ref. \cite{adrowe1} that those modes contribute to the creation of thick spectral barriers. The formation of spectral walls in the dynamics of kinks has also been studied in models containing two scalar fields \cite{adam.2021}.

A new family of multifield models has also been recently explored \cite{blm}. In these models, the kinetic term was modified. This modification leads to the existence of intriguing internal structures that mimic a two-kink profile. Currently, such a profile is understood to be a direct manifestation of geometric constrictions, see Ref. \cite{jab}. The presence of such constrictions can therefore be simulated via adjustments on the kinetic terms, see Refs. \cite{blm,balbazmar}. In Ref. \cite{bazmarmen}, the authors have explored the appearance of compact and long-range solutions through geometrically constrained kinklike configurations. More recently, this mechanism has been extended to models with three scalar fields, see Ref. \cite{marmen}.

Another relevant investigation concerning geometrical constrictions is presented in Ref. \cite{hiatma}. There, magnetic structures inherent to a Ni$_{81}$Fe$_{19}$ thin film were considered. These structures exhibit domain wall nucleation in nanoscale constrictions. Moreover, Ref. \cite{clahu} reported the fabrication of nanostructures designed to host constrained walls. Also, research of confined walls within magnetic nanotubes revealed that their magnetization structure has a two-kink behavior depending on the size of the tubes, see Ref. \cite{chego}.

Geometrical constrictions can also play a crucial role in the formation of domain walls in kink dynamics in models with two fields. At the nanoscale, for instance, electric charges flow as a current with two-kink properties \cite{thivan}. Moreover, it is currently known that geometric modifications influence the behavior of fermions \cite{fermion}.

Very recently, one of us has studied the influence of geometrical constrictions on the dynamics of colliding kinks \cite{joao}. The model considered has two interacting real scalar fields, i.e. $\phi$ and $\chi$. They self-interact through a potential that is basically a sum of the standard $\phi^4$ with the $\chi^4$ one. The kinetic term for $\phi$ has a nonstandard form defined in terms of a coupling function that depends only on $\chi$.

Motivated by that research, we now go further and study a scenario in which the dynamics of one of the fields is commanded by the sine-Gordon potential. The sine-Gordon model is a well-known example of an integrable theory, from which no radiation is emitted during a typical kink-antikink scattering. However, now the sine-Gordon field $\phi$ interacts with a non-integrable one submitted to the standard $\chi^4$ potential. As a consequence, given the non-integrability of the composite scenario, some radiation is expected to be emitted from the collision process. In this case, it is interesting to consider some properties of the overall model. For instance, whether the geometrically constrained sine-Gordon kink develops internal structures that mimic a two-kink profile, and how those structures influence the scattering process itself. It is also interesting to verify whether the emission of radiation can be modulated by the value of a real parameter that enters the definition of the generalizing function
f($\chi$). We also discuss the conditions to be satisfied by a {\textit{physically acceptable}} interaction between the fields.

This manuscript is organized as follows: in Section \ref{secII}, we introduce our model, its lagrangian density and
minimize the total energy by using the Bogomol'nyi-Prasad-Sommerfield method. The Bogomol'nyi bound for the energy is obtained
and the first-order equations of motion are derived. In Section \ref{secIII}, we particularize our investigation. We take the standard sine-Gordon potential for
the field $\phi$ and the canonical $\chi^4$ for $\chi$. We also set $f(\chi)$. We obtain the corresponding BPS kinks analytically and investigate their stability against small perturbations. We then study numerically the kink-antikink scattering for different values of the parameters of the model and discuss its most important results. Finally, Section \ref{secIV} brings our conclusions and perspectives regarding future research.

\section{The general model and its BPS structure} \label{secII}

We consider a $(1+1)$-dimensional model with two scalar fields, i.e. $\chi(x,t)$ and $\phi(x,t)$. The potential $V\left( \chi ,\phi \right)$ defines the vacuum structure of the theory. The corresponding lagrangian density is given by
\begin{equation}
\mathcal{L}=\frac{1}{2}f\left( \chi \right) \partial _{\mu }\phi
\partial ^{\mu }\phi +\frac{1}{2}\partial _{\mu }\chi \partial ^{\mu }\chi
-V\left( \chi ,\phi \right) \text{.} \label{lr1}
\end{equation}%
Note the presence of the nontrivial function $f\left( \chi \right)$. This function couples the fields in a non-usual way. As a result, it mimics the effects of geometrical constrictions now applied on $\phi$. Here, we are working in a Minkowski spacetime with a $(+-)$ signature. Also, we take all the fields, coordinates and coupling constants dimensionless [after being mass rescaled].

We choose for the potential
\begin{equation}
V\left( \chi ,\phi \right) =\frac{1}{2}\frac{W_{\phi }^{2}}{f\left( \chi
\right) }+\frac{1}{2}W_{\chi }^{2}\text{,}\label{p}
\end{equation}%
with $W_{\phi}=\partial_{\phi} W$ and $W_{\chi}=\partial_{\chi} W$. Here, $W=W\left( \chi ,\phi \right)$ is the {\it{superpotential}}. Eq. (\ref{p}) guarantees the existence of first-order equations, see Ref. \cite{blm}.

We intend to study the BPS soliton solutions and their scattering. By this way, one needs first to write the Euler-Lagrange equations of motion. For the lagrangian density (\ref{lr1}), these equations read
\begin{equation}
f\left( \frac{\partial ^{2}\phi }{\partial t^{2}}-\frac{\partial ^{2}\phi }{%
\partial x^{2}}\right) +\frac{df}{d\chi }\left( \frac{\partial \chi }{%
\partial t}\frac{\partial \phi }{\partial t}-\frac{\partial \chi }{\partial x%
}\frac{\partial \phi }{\partial x}\right) =-\frac{\partial V}{\partial \phi }%
 \text{,}  \label{emplr1}
\end{equation}%
\begin{equation}
\frac{\partial ^{2}\chi }{\partial t^{2}}-\frac{\partial ^{2}\chi }{\partial
x^{2}}+\frac{1}{2}\frac{df}{d\chi }\left[ \left( \frac{\partial \phi }{%
\partial x}\right) ^{2}-\left( \frac{\partial \phi }{\partial t}\right) ^{2}%
\right] =-\frac{\partial V}{\partial \chi }\text{,}
\label{emclr1}
\end{equation}
with the potential given in (\ref{p}).

We now proceed to the minimization of the total energy of model. We then apply the well-known BPS method \cite{prasom,bogo}. We begin by writing the energy density for static fields
\begin{equation}
\varepsilon =\frac{1}{2}f\left( \chi \right) \left( \frac{%
d\phi }{dx}\right) ^{2}+\frac{1}{2}\left( \frac{d\chi }{dx}\right) ^{2}+%
\frac{1}{2}\frac{W_{\phi }^{2}}{f\left( \chi \right) }+\frac{1}{2}W_{\chi
}^{2}\text{,}
\end{equation}%
as a sum of squared terms, i.e.,
\begin{equation}
\varepsilon =\frac{f}{2}\left( \frac{d\phi }{dx}\mp \frac{W_{\phi }}{f}%
\right) ^{2}+\frac{1}{2}\left( \frac{d\chi }{dx}\mp W_{\chi }\right) ^{2}\pm
\frac{dW}{dx}\text{.}
\label{sumsquare}
\end{equation}%

In the expression above, we assume that the squared terms vanish. As a consequence, we get the first-order BPS equations for $\phi(x)$ and $\chi(x)$
\begin{equation}
\frac{d\phi }{dx}=\pm 
\frac{W_{\phi }}{f}\text{,}\label{bpsphi}
\end{equation}%
\begin{equation}
\frac{d\chi }{dx}=\pm W_{\chi }\text{,}%
\label{bpschi}
\end{equation}%
whose solutions minimize the energy related to (\ref{sumsquare}), as desired.

The total energy of the BPS configurations then saturates to
\begin{equation}
E=E_{BPS}=\int \varepsilon _{BPS}dx=\pm \Delta W\text{,}
\label{ebps}
\end{equation}%
where $E_{BPS}$ stands for the so-called Bogomol'nyi bound. Here, we have defined
\begin{equation}
\Delta W=W\left( x\rightarrow +\infty \right) -W\left( x\rightarrow -\infty
\right)\text{,}
\end{equation}
while
\begin{equation}
\varepsilon _{BPS}=\pm {\frac{dW }{dx}} \text{,}
\label{edbps}
\end{equation}%
is the energy density of the first-order profiles.

The BPS Eq. (\ref{bpschi}) does not depend on $\phi$. Therefore, it can be solved separately once $W_{\chi}(\chi,\phi)$ is known.

On the other hand, the solution for $\phi$ depends on both $W_{\phi}(\chi,\phi)$ and $f(\chi)$. As a consequence of such a dependence, the field $\phi(x)$ can be ``trapped" by $\chi(x)$. That field can then develop a double-kink profile. Such a profile is currently known to be caused by geometrical constrictions.

This is a direct consequence of the interaction between $\phi$ and $\chi$ as mediated by $f(\chi)$. It sounds then interesting to consider how these fields mutually interact. As we discuss above, given a well-behaved superpotential $W(\chi,\phi)$, it is always possible to find a BPS solution for $\chi(x)$. However, in view of Eq. (\ref{bpsphi}), a real field $\phi(x)$ can only be obtained for some particular $f(\chi)$. In other words, those physically acceptable choices for $f$ are restricted. Naturally, this limitation also applies to the way $\phi$ and $\chi$ interact. Regarding this, the present study may provide important insights about this issue.

In Ref. \cite{joao}, the authors have studied BPS solutions (including their scattering) in an enlarged scenario with a nontrivial $f(\chi)$. In that case, the superpotential was chosen as the sum between the standard $\phi^4$ and $\chi^4$ ones. The authors have verified that both fields assume a kinklike profile. Also, the BPS solution for $\phi$ develops internal structures and a two-kink shape. The interested reader is referred to Ref. \cite{joao} for further details.

We now investigate the effects of geometrical constrictions when $W(\chi,\phi)$ includes the $\phi$ sine-Gordon superpotential. Here, we keep the $\chi^4$ term. We then focus on those effects applied to the sine-Gordon kink. In particular, whether this kink develops internal structures that can be associated to a two-kink profile.

It is interesting to highlight that our proposal involves an integrable field. In this way, when the coupling between $\phi$ and $\chi$ decreases, these fields behave {\textit{almost}} independently. A collision in the integrable sine-Gordon sector $\phi$ is then expected to emit almost no radiation. On the other hand, when the interaction gets stronger, geometrical constrictions play an important role. In this case, radiation is emitted in both sectors. These observations suggest that the coupling mediated by $f(\chi)$ can be used to modulate the radiation that escapes during the scattering. This possibility is also discussed below.

\section{sine-Gordon field under geometrical constrictions} \label{secIII}

We assume that the field $\phi$ self-interacts according to the sine-Gordon potential, while $\chi$ is driven by the $\chi^4$ one. We then study the effects caused by the mutual interaction between these fields. In particular, we explore the conditions under which $\phi$ is trapped by $\chi$. With this aim in mind, we choose the superpotential as
\begin{equation}
W\left( \phi ,\chi \right) =-\eta \cos \left[ \phi \right] +\alpha \chi
\left( 1-\frac{1}{3}\chi ^{2}\right) \text{,}
\label{w0}
\end{equation}
via which Eq. (\ref{p}) leads to%
\begin{equation}
V\left( \chi ,\phi \right) =%
\frac{\eta ^{2}}{2f\left( \chi \right) }\sin ^{2}\left[ \phi \right] +\frac{1%
}{2}\alpha ^{2}\left( 1-\chi ^{2}\right) ^{2}\text{.}
\label{px0}
\end{equation}%
Here, $\eta$ and $\alpha$ are positive constants that control the masses of $\phi$ and $\chi$, respectively.

For the sake of simplicity, we choose $f\left( \chi \right)$ as used in Ref. \cite{joao}, i.e.
\begin{equation}
f\left( \chi \right) =\frac{1+\lambda }{1+\lambda \chi ^{2}}\text{,}
\label{fx100a1}
\end{equation}%
where $\lambda$ stands for a positive parameter. Once $\chi$ interpolates between $-1$ to $+1$, one gets from Eq. (\ref{fx100a1}) a regular profile without divergences. Moreover, the limit $\lambda \rightarrow 0$ leads to $f(\chi) \rightarrow 1$. In this regime, the two sectors decouple, and the fields can be treated individually. The mutual interaction is weak, and the effects of geometrical constrictions are small. On the other hand, when $\lambda \rightarrow \infty$, $f(\chi)$ behaves as $\chi^{-2}$. In this case, $\phi$ and $\chi$ are strongly coupled. As a consequence, geometrical constrictions have a great influence, and relevant effects appear. In this manuscript, we therefore consider some interesting values of $\lambda$.

\begin{figure*}[!ht]
\begin{center}
  \centering
    \includegraphics[{angle=0,width=9cm,height=6.2cm}]{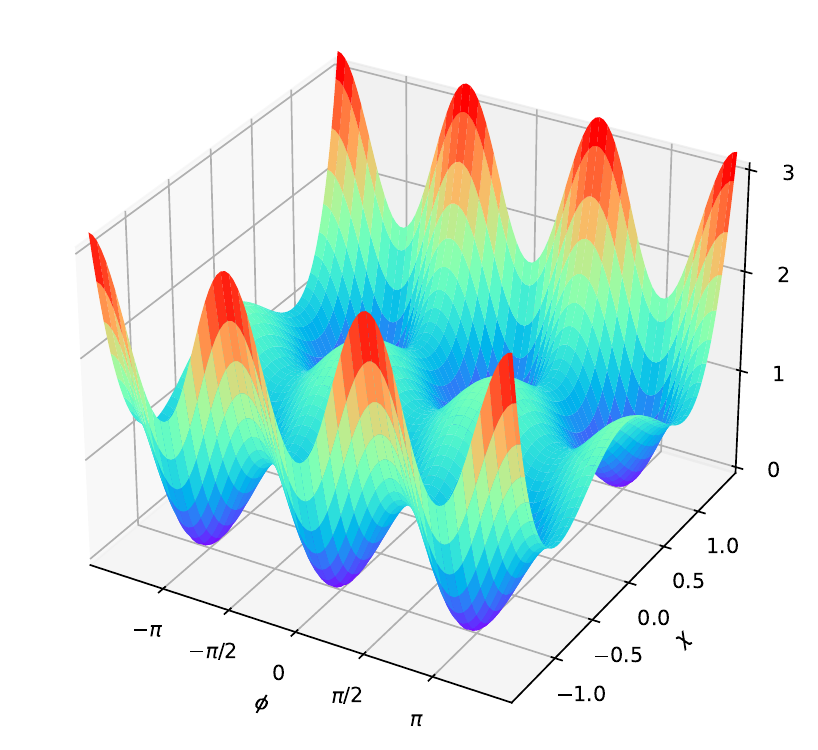}\label{tAs10_01}
    \includegraphics[{angle=0,width=7cm,height=6cm}]{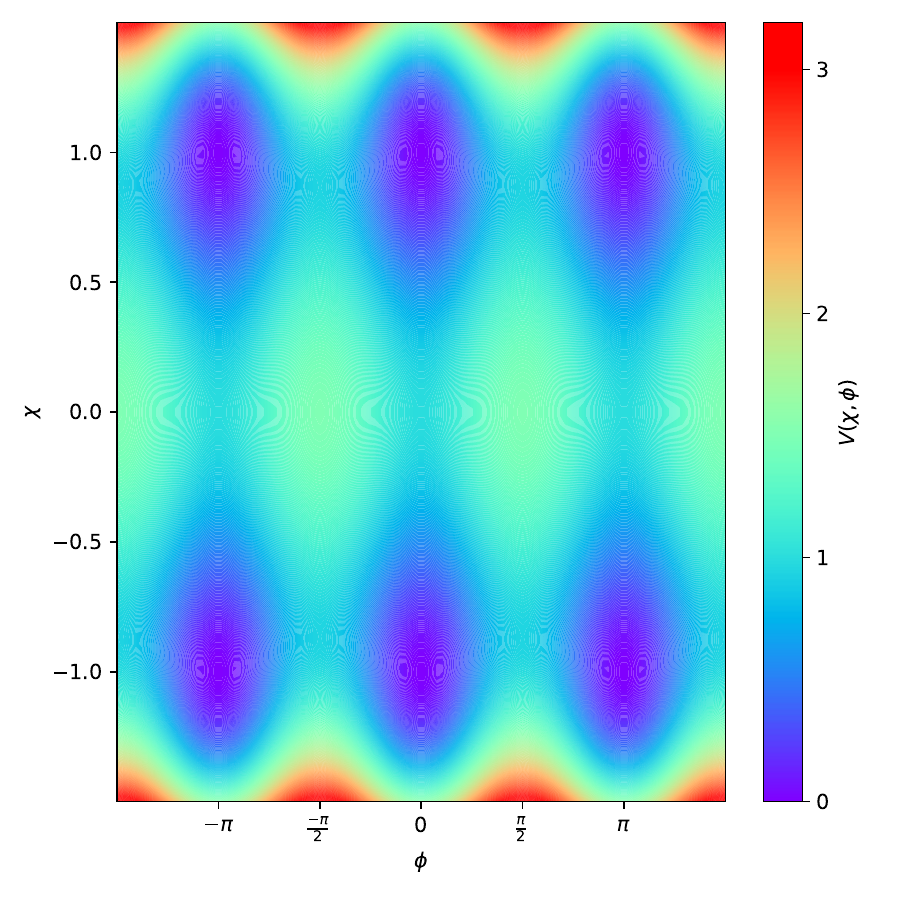}\label{tAs10_20}
    \vspace{-0.5cm}
  \caption{Potential (\ref{p00}) for $\eta=\sqrt{2}$, $\alpha=\sqrt{2}$ and $\lambda=1$. Left (right): 3-dimensional (top) view. The vacuum manifold presents 6 different minima (highlighted in purple).}
  \label{times_A_sigma_1000}
\end{center}
\end{figure*}

Explicitly, Eqs. (\ref{px0}) and (\ref{fx100a1}) lead to the effective potential%
\begin{equation}
V\left( \chi ,\phi \right) =%
\frac{1+\lambda \chi ^{2}}{2\left( 1+\lambda \right) }\eta ^{2}\sin ^{2}%
\left[ \phi \right] +\frac{1}{2}\alpha ^{2}\left( 1-\chi ^{2}\right) ^{2}%
\text{.}\label{p00}
\end{equation}
This potential is plotted in Fig. \ref{times_A_sigma_1000} for $-\frac{3\pi}{2}<\phi(x)<\frac{3\pi}{2}$ and $-\frac{3}{2}<\chi(x)<\frac{3}{2}$. Those plots reveal the existence of 6 minima and thus 11 topological sectors (i.e. pairs of adjacent minima). These sectors support the occurrence of 22 families of topological solutions. These solutions are mutually related via the $Z_2 \times Z_2$ symmetry inherent to the model (\ref{lr1}). Specifically, there are 8 symmetrically equivalent sine-Gordon solutions for $\phi$ with trivial $\chi$. In addition, we can identify the existence of six $\chi^4$-kinks with trivial $\phi$. There are also 8 families of solutions with nontrivial $\phi$ and $\chi$.

In this work, we focus on this last family of solutions. This is because, as we demonstrate, this family engenders BPS configurations with internal structures. These solutions then mimic a two-kink profile that is known to be related to geometric constrictions.

{\subsection{BPS solutions}}

We now study the first-order configurations themselves. With this aim, we consider the BPS Eqs. (\ref{bpsphi}) and (\ref{bpschi}). We observe that, given the potential (\ref{p00}), a localized kink must satisfy
\begin{equation}
\phi _{k}\left( x\rightarrow -\infty \right) \rightarrow 0\text{ \ \ and \ \ 
}\phi _{k}\left( x\rightarrow +\infty \right) \rightarrow \pi \text{,}
\label{bcphik001a1}
\end{equation}%
\begin{equation}
\chi _{k}\left( x\rightarrow -\infty \right) \rightarrow -1\text{ \ \ and \
\ }\chi _{k}\left( x\rightarrow +\infty \right) \rightarrow +1\text{.}
\label{bcchik00a1}
\end{equation}%
It is interesting to note that the conditions above are the same as the standard ones when the fields $\phi$ 
and $\chi$ are decoupled. In other words, a sine-Gordon kink and the field $\chi$ behave asymptotically as in Eqs. (\ref{bcphik001a1}) and (\ref{bcchik00a1}) for the standard sine-Gordon and $\chi^4$ models respectivelly.
Therefore, the interaction mediated by $f(\chi)$ (see Eq. (\ref{fx100a1})) does not affect the asymptotic values of the fields at the boundaries. As a consequence, the effects due to geometric constrictions are expected to appear around the center of the BPS solutions. Eventually, other choices for $f(\chi)$ may change the asymptotic field values. However, this discussion is beyond the scope of this manuscript.

Given the conditions (\ref{bcphik001a1}) and (\ref{bcchik00a1}), we calculate%
\begin{equation}
\Delta W_{k}=2\left( \eta +\frac{2}{3}\alpha \right)\text{,}
\end{equation}
that leads to
\begin{equation}
E_{BPS,k}=+\Delta W_{k}=2\left( \eta +\frac{2}{3}\alpha \right) \text{.}\label{e00}
\end{equation}
This result is the total energy of the BPS kink, see Eq. (\ref{ebps}).

As one can see, the above result does not depend on $f(\chi)$. In other words, the interaction between $\phi$ and $\chi$ does not affect the energy of the BPS configuration - in particular, even when that coupling is strong, i.e. $\lambda$ is large.

\begin{figure*}[!ht]
\begin{center}
  \centering
    {\includegraphics[width=1.0  \textwidth]{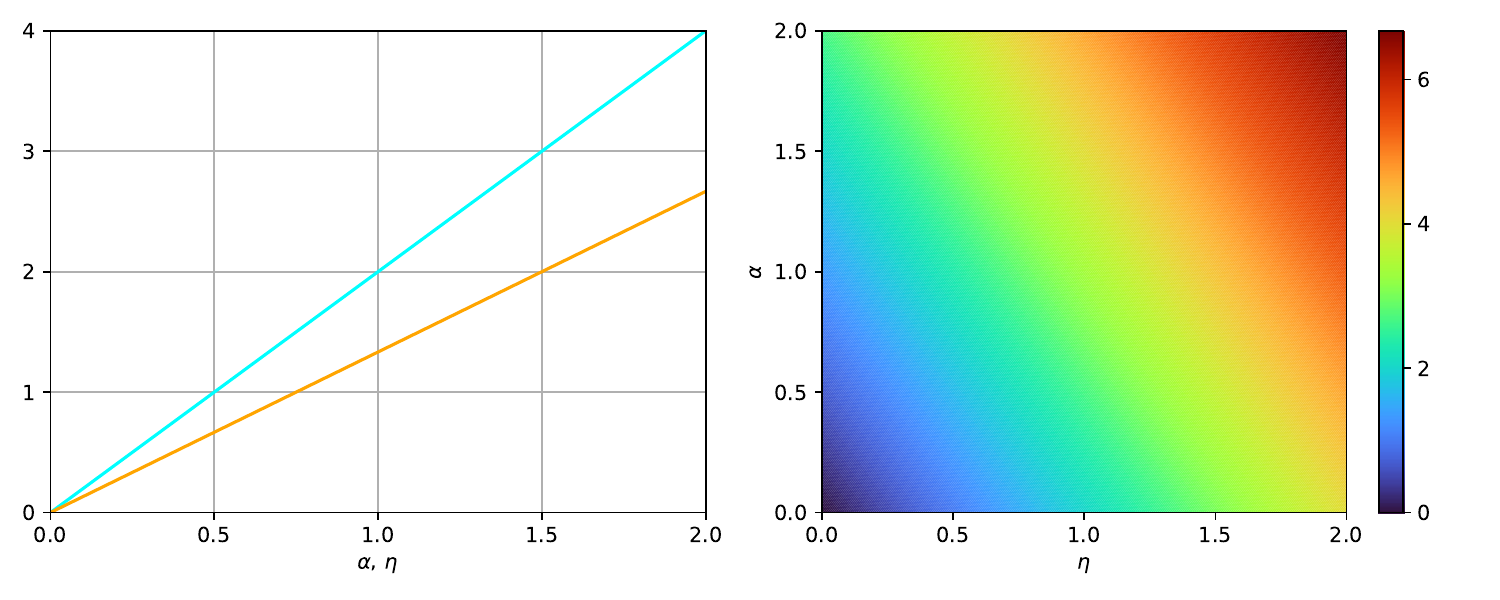}}\label{tAs10_010}
    \vspace{-1.5cm}
  \caption{Left: individual masses of $\phi$ (cyan) and $\chi$ (orange) as functions of $\eta$ and $\alpha$, respectively. Right: total mass Eq. (\ref{e00}) in terms of both $\eta$ and $\alpha$. These results do not depend on $f(\chi)$.}
  \label{f_energy}
\end{center}
\end{figure*}

The value for BPS energy given in Eq. (\ref{e00}) can also be seen as the total mass of the composite BPS solution. Therefore, its first and second contributions can be considered as the individual masses of $\phi$ and $\chi$, respectively. These masses are plotted in Fig. (\ref{f_energy}) (left) as functions of $\eta$ and $\alpha$. It is possible to compare them for different values of those parameters. The total mass also appears in the same Fig. (\ref{f_energy}) (right) as a simultaneous function of both $\eta$ and $\alpha$.

Using Eq. (\ref{bcchik00a1}), the first-order Eq. (\ref{bpschi}) provides the kink solution
\begin{equation}
\chi _{k}\left( x\right) =\tanh \left[ \alpha \left( x-x_{0}\right) \right] \text{.}
\label{schik001a1}
\end{equation}
Here, $\alpha^{-1}$ is the width of the kink, while $x_{0}$ stands for its position.

In what follows, we consider the BPS solution for $\phi(x)$. We first rewrite Eq. (\ref{bpsphi}) as%
\begin{equation}
\frac{d\phi }{dy}=\pm \eta \sin \left[ \phi \right]\text{.}
\label{ephiy0a10}
\end{equation}%
In this case, we have introduced a new spatial coordinate $y$. Its dependence on $x$ is obtained as the solution of
\begin{equation}
\frac{dy}{dx}=f^{-1}(\chi(x))\text{,}  \label{fc0a1}
\end{equation}%
where $\chi(x)$ is given by Eq. (\ref{schik001a1}). In summary, $\phi$ depends on $y$ and thus it depends on $\chi(x)$ and $f(\chi)$ (i.e. the way the field sectors interact). This dependence is discussed in more detail below.

The kink solution of Eq. (\ref{ephiy0a10}) then reads
\begin{equation}
\phi _{k}\left( y\right) =2\arctan \left[ \exp \left( \eta \left(
y-y_{0}\right) \right) \right] \text{.}  \label{phikya1}
\end{equation}%

Now, it is time to rewrite the solution above as a function of $x$. It is then necessary to solve Eq. (\ref{fc0a1}) itself. Obviously the solution depends on the expressions for $f$ and $\chi$. In the present manuscript, these expressions are given by Eq. (\ref{fx100a1}) and Eq. (\ref{schik001a1}), respectively. So, Eq. (\ref{fc0a1}) becomes
\begin{equation}
\frac{dy}{dx}=\frac{1+\lambda \tanh ^{2}\left[ \alpha \left( x-x_{0}\right) %
\right] }{1+\lambda }\text{,}  \label{dydxa1}
\end{equation}%
whose solution is
\begin{equation}
y(x)=\left( x-x_{0}\right) -\frac{\lambda }{\alpha \left( 1+\lambda \right) }%
\tanh \left[ \alpha \left( x-x_{0}\right) \right] \text{.}  \label{yxa1}
\end{equation}%

Therefore, Eq. (\ref{phikya1}) can be finally written as%
\begin{equation}
\phi _{k}\left( x\right)
=2\arctan \left[ \exp \left( \eta \left( y_{0}+\left( x-x_{0}\right) -\frac{%
\lambda }{\alpha \left( 1+\lambda \right) }\tanh \left[ \alpha \left(
x-x_{0}\right) \right] \right) \right) \right] \text{.}  \label{sphik001a1}
\end{equation}%
It represents a kink profile that satisfies the conditions in Eq. (\ref{bcphik001a1}).

It is now clear how the coupling mediated by $f(\chi)$ affects the BPS field $\phi_{k}(x)$. For a well-behaved superpotential $W(\chi,\phi)$, it is always possible to solve Eq. (\ref{bpschi}) and find a BPS kink $\chi_{k}(x)$. In addition, one can always rewrite Eq. (\ref{bpsphi}) in terms of the coordinate $y$. Then, a BPS solution $\phi_{k}(y)$ emerges. However, the relation between $y$ and $x$ is given as the solution to Eq. (\ref{fc0a1}). This means that it depends on both $\chi_{k}(x)$ and $f(\chi)$. Naturally, a {\textit{real}} relation $y(x)$ can only be obtained for some particular $f$. The choices for $f$ are then restricted which means that way the fields interact is also restricted.

We now return to our specific case. The antikinks can be obtained via the $Z_2 \times Z_2$ symmetry inherent to Eq. (\ref{lr1}). They read
\begin{equation}
\chi _{ak}\left( x\right) =-\tanh \left[ \alpha \left(
x-x_{0}\right) \right] \text{,}  \label{schiak001a1}
\end{equation}%
\begin{equation}
\phi _{ak}\left( x\right) =2\arctan \left[ \exp \left( -\eta \left(
y_{0}+\left( x-x_{0}\right) -\frac{\lambda }{\alpha \left( 1+\lambda \right) 
}\tanh \left[ \alpha \left( x-x_{0}\right) \right] \right) \right) \right] 
\text{.}  \label{sphiak001a1}
\end{equation}%
These solutions satisfy Eq. (\ref{bcphik001a1}) and Eq. (\ref{bcchik00a1}) in the opposite spatial direction, i.e. $\chi _{ak}\left( x\rightarrow \mp \infty \right) \rightarrow \pm 1$, with $\phi _{ak}\left( x\rightarrow -\infty \right) \rightarrow \pi$ and $\phi _{ak}\left( x\rightarrow +\infty \right) \rightarrow 0$.

\begin{figure*}[!ht]
\begin{center}
  \centering
    \includegraphics[width=1.0 \textwidth]{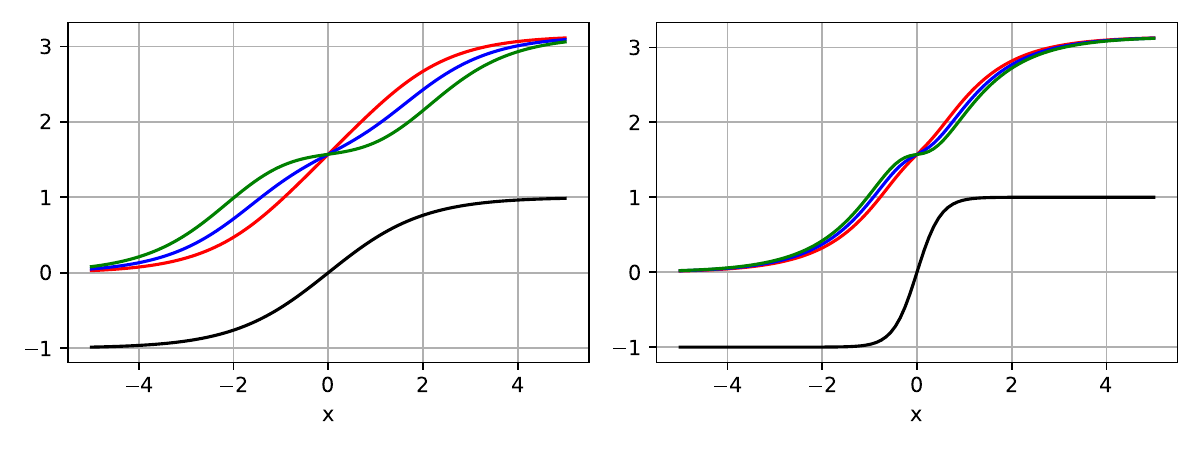}\label{tAs10_1}
    \vspace{-1.5cm}
  \caption{BPS kink profiles $\chi_{k}(x)$ (black line) and $\phi_{k}(x)$ (coloured lines) with $\lambda=0.6$ (red), $\lambda=2$ (blue) and $\lambda=10$ (green). Here, $\alpha=0.5$ (right) and $\alpha=2.0$ (left). We have fixed $y_0=0$, $x_0=0$ and $\eta=1$.}
  \label{times_A_sigma_100}
\end{center}
\end{figure*}

In Fig. \ref{times_A_sigma_100}, we plotted $\chi_k(x)$ and $\phi_k(x)$ for different $\alpha$ and $\lambda$. Here, $y_0$, $x_0$ and $\eta$ are fixed. As $\alpha$ increases, the fields become more localized around the origin. In this case the field
$\chi_k(x)$ ``traps" the field $\phi_k(x)$. In addition, for small $\lambda$, the coupling between $\chi$ and $\phi$ is weak. Then, the effects of geometrical constrictions are small. As a consequence, the solutions are almost standard. On the other hand, as $\lambda$ increases, the interaction gets stronger, and the constrictions start to play a relevant role. In this case, $\phi_k(x)$ develops an internal structure around $x=0$. The resulting solution then mimics a two-kink profile. Note 
that the profile of the field $\chi_k(x)$ is not affected.

Now taking into account Eq. (\ref{bpsphi}) and Eq. (\ref{bpschi}), the energy density given in Eq. (\ref{edbps}) can be written as
\begin{equation}
\varepsilon _{BPS}=\varepsilon _{\phi}+\varepsilon _{\chi}\text{,}
\end{equation}%
where
\begin{equation}
\varepsilon _{\phi}=\frac{W_{\phi }^{2}}{%
f\left( \chi \right) }=\frac{\eta ^{2}\left( 1+\lambda \chi
^{2}\right) }{1+\lambda }\sin ^{2}\left[ \phi \right] \text{,}
\end{equation}%
\begin{equation}
\varepsilon _{\chi}=W_{\chi }^{2}=\alpha ^{2}\left(
1-\chi ^{2}\right) ^{2}\text{.}
\end{equation}%
They represent, respectively, the energy densities related to the fields $\phi$  and $\chi$.

\begin{figure*}[!ht]
\begin{center}
  \centering
    \includegraphics[width=1.0 \textwidth]{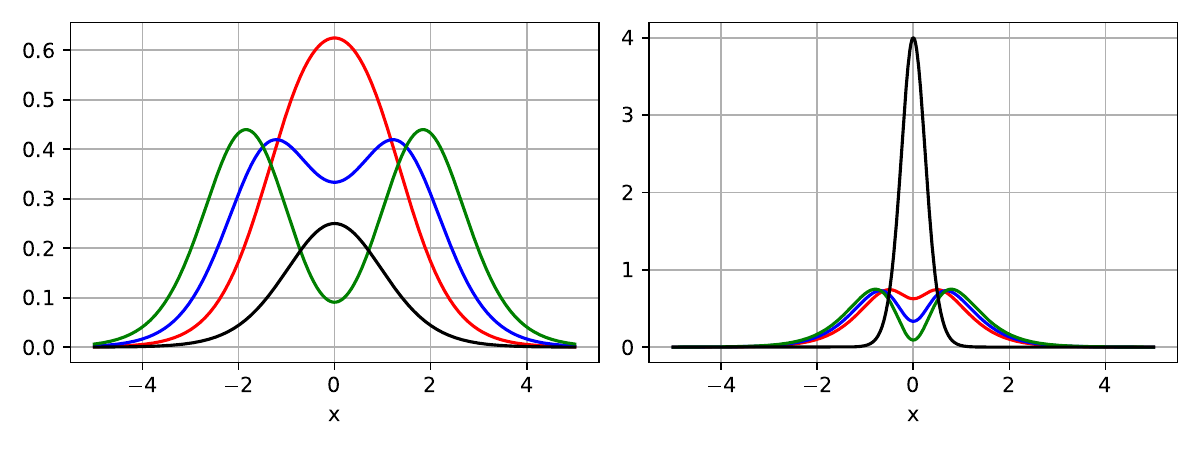}\label{fig4a}
    \vspace{-1.5cm}
  \caption{Energy densities $\varepsilon_{\chi}(x)$ (black line) and $\varepsilon_{\phi}(x)$ (coloured lines). Conventions as in Fig. 3.}
  \label{fig4}
\end{center}
\end{figure*}

Fig. \ref{fig4} shows $\varepsilon _{\phi}(x)$ and $\varepsilon _{\chi}(x)$ for different $\alpha$ and $\lambda$. Again, $y_0$, $x_0$ and $\eta$ are fixed. We have used Eqs. (\ref{schik001a1}) and (\ref{sphik001a1}) for $\chi_k(x)$ and $\phi_k(x)$, respectively. The densities become more localized as $\alpha$ increases. Again, for small $\lambda$, the solutions behave as for the standard sine-Gordon model. As $\lambda$ increases, $\varepsilon _{\phi}$ develops a two-lump profile. This is a consequence of the two-kink profile solution plotted in Fig. \ref{times_A_sigma_100}.

In what follows, we study the linear stability of the BPS kinks introduced above. We intend to verify the existence of internal modes. The energy exchange between these modes may explain the occurrence of multiple-bounce windows during the collision process.


{\subsection{Spectrum of small excitations}}

We now investigate the stability of Eqs. (\ref{schik001a1}) and (%
\ref{sphik001a1}) against small perturbations. Regarding this, we look for internal modes that play a crucial role during the scattering process. Note that a similar study is presented in Refs. \cite{barisa,banari}.

We write
\begin{equation}
\phi \left( x,t\right) =\phi _{k}\left( x\right) +\mu \left( x\right) \cos
\left( \omega t\right)  \label{ls01a}
\end{equation}%
\begin{equation}
\chi \left( x,t\right) =\chi _{k}\left( x\right) +\psi \left( x\right) \cos
\left( \omega t\right) \label{ls02a}
\end{equation}%
where $\mu(x)$ and $\psi(x)$ are small perturbations around the static BPS kinks and whose profiles we want to determine.

Up to the first order in $\mu$ and $\psi$, Eqs. (\ref{emplr1}) and (\ref{emclr1}) become
\begin{equation}
\widehat{f}\frac{d^{2}\mu }{dx^{2}}+\widehat{f}_{\chi } \left(\frac{d\chi _{k}}{dx}%
\frac{d\mu }{dx}+\frac{d\phi _{k}}{dx}\frac{d\psi }{dx}\right)-%
\widehat{V}_{\phi \phi }\mu -\left( \widehat{V}_{\chi \phi }-\widehat{f}%
_{\chi \chi }\frac{d\phi _{k}}{dx}\frac{d\chi _{k}}{dx}-\widehat{f}_{\chi }%
\frac{d^{2}\phi _{k}}{dx^{2}}\right) \psi =-\omega ^{2}\widehat{f}\mu\text{,}
\label{le1a}
\end{equation}%
\begin{equation}
\frac{d^{2}\psi }{dx^{2}}-\widehat{f}_{\chi }\frac{d\phi _{k}}{dx}\frac{%
d\mu }{dx}-\widehat{V}_{\phi \chi }\mu -\left( \widehat{V}_{\chi \chi }+%
\frac{1}{2}\widehat{f}_{\chi \chi }\left( \frac{d\phi _{k}}{dx}\right)
^{2}\right) \psi =-\omega ^{2}\psi \text{,}  \label{le2a}
\end{equation}%
where the symbol `` \hspace{0.05cm}$\widehat{}$ \hspace{0.05cm}" above a function
means that it must be evaluated at the BPS profile.

In general, Eqs. (\ref{le1a}) and (\ref{le2a}) do not admit analytical solutions. Also, due to the their intricate structure, solving them numerically is a quite hard task. However, there are few particular cases of interest where analytical solutions can be found. We explore them below.

\textit{1. Zero modes}. We first consider the translational modes. Regarding this, we choose $\omega =0$. Then, Eqs. (\ref{le1a}) and
(\ref{le2a}) can be written as%
\begin{eqnarray}
&&\frac{d^{2}\mu }{dx^{2}}-2\alpha \lambda \frac{\chi _{k}\left( 1-\chi
_{k}^{2}\right) }{1+\lambda \chi _{k}^{2}}\frac{d\mu }{dx}-\frac{2\eta
\lambda \chi _{k}}{1+\lambda }\sin \left[ \phi _{k}\right] \frac{d\psi }{dx}-%
\frac{\eta ^{2}\left( 1+\lambda \chi _{k}^{2}\right) ^{2}}{\left( 1+\lambda
\right) ^{2}}\cos \left[ 2\phi _{k}\right] \mu   \notag \\
&&=\frac{2\eta \lambda }{1+\lambda }\left( \frac{2\eta }{1+\lambda }\chi
_{k}\cos \left[ \phi _{k}\right] +\frac{\alpha \left( 1-\lambda \chi
_{k}^{2}\right) }{\left( 1+\lambda \chi _{k}^{2}\right) ^{2}}\left( 1-\chi
_{k}^{2}\right) \right) \left( 1+\lambda \chi _{k}^{2}\right) \sin \left[
\phi _{k}\right] \psi \text{,}  \label{leaaxx0}
\end{eqnarray}%
\begin{equation}
\frac{d^{2}\psi }{dx^{2}}+2\eta \lambda \frac{\chi _{k}\sin \left[ \phi _{k}%
\right] }{1+\lambda \chi _{k}^{2}}\frac{d\mu }{dx}-\frac{\eta ^{2}\lambda }{%
1+\lambda }\chi _{k}\sin \left[ 2\phi _{k}\right] \mu =2\left[ \frac{2\eta
^{2}\lambda ^{2}}{1+\lambda }\frac{\chi _{k}^{2}\sin ^{2}\left[ \phi _{k}%
\right] }{1+\lambda \chi _{k}^{2}}-\alpha ^{2}\left( 1-3\chi _{k}^{2}\right) %
\right] \psi \text{,}  \label{lebbxx0}
\end{equation}%
where we have used Eqs. (\ref{bpsphi}) and (\ref{bpschi}).
We have also used Eqs. (\ref{fx100a1}) and (\ref{p00}) for $f(\chi)$ and $V(\chi,\phi)$.
In the above expressions, $\chi _{k}\left( x\right) $ and $\phi _{k}(x)$ are given 
by Eqs. (\ref%
{schik001a1}) and (\ref{sphik001a1}), respectively.

First, note that Eqs. (\ref{leaaxx0}) and (\ref{lebbxx0}) admit the solution%
\begin{equation}
\mu_{0} \left( x\right) =\sin \left[ 2\arctan \left[ \exp \left( \eta \left(
y_{0}+\left( x-x_{0}\right) -\frac{\lambda }{\alpha \left( 1+\lambda \right) 
}\tanh \left[ \alpha \left( x-x_{0}\right) \right] \right) \right) \right] %
\right]\text{,}
\end{equation}%
with $\psi_{0} (x)=0$. This corresponds to the translation mode for $\phi$.

Second, we highlight that Eqs. (\ref{leaaxx0}) and (\ref{lebbxx0}) are also solved by%
\begin{equation}
\psi_{0} \left( x\right) =\text{sech}%
^{2}\left[ \alpha \left( x-x_{0}\right) \right] \text{,}
\end{equation}%
\begin{eqnarray}
\mu_{0} \left( x\right)  =&&\frac{1+\lambda \tanh ^{2}\left[ \alpha \left(
x-x_{0}\right) \right] }{1+\lambda } \times   \notag \\
&&\sin \left[ 2\arctan \left[ \exp \left(
\eta \left( y_{0}+\left( x-x_{0}\right) -\frac{\lambda }{\alpha \left(
1+\lambda \right) }\tanh \left[ \alpha \left( x-x_{0}\right) \right] \right)
\right) \right] \right]\text{.}
\end{eqnarray}%
This set represents the translation of both $\phi$ and $\chi$.


\textit{2. The case with trivial }$\chi $. Whether we choose $%
\chi =\pm 1$, one gets that $\widehat{f} =1$. The first-order Eq. (\ref{bpschi}) is then identically satisfied. In addition, Eq. (\ref{bpsphi}) provides
\begin{equation}
\phi _{k}\left( x\right) =2\arctan \left[ \exp \left( \eta \left(
x-x_{0}\right) \right) \right] \text{,}  \label{sgs}
\end{equation}%
i.e. the canonical BPS sine-Gordon kink.

The linearized Eqs. (\ref{le1a}) and (\ref{le2a}) then become
\begin{equation}
\frac{d^{2}\mu }{dx^{2}}\mp \frac{2\lambda \eta }{1+\lambda }\sin \left[
\phi _{k}\right] \frac{d\psi }{dx}-\eta ^{2}\cos \left[ 2\phi _{k}\right]
\mu \mp \frac{2\eta ^{2}\lambda }{1+\lambda }\sin \left[ 2\phi _{k}\right]
\psi =-\omega ^{2}\mu \text{,}  \label{eq1aa}
\end{equation}%
\begin{equation}
\frac{d^{2}\psi }{dx^{2}}\pm \frac{2\lambda \eta }{1+\lambda }\sin \left[
\phi _{k}\right] \frac{d\mu }{dx}\mp \frac{\eta ^{2}\lambda }{1+\lambda }%
\sin \left[ 2\phi _{k}\right] \mu -4\left( \alpha ^{2}+\frac{\eta
^{2}\lambda ^{2}}{\left( 1+\lambda \right) ^{2}}\sin ^{2}\left[ \phi _{k}%
\right] \right) \psi =-\omega ^{2}\psi \text{,}  \label{eq2aa}
\end{equation}%
where $\phi _{k}$ is now given by (\ref{sgs}).

Equations (\ref{eq1aa}) and (\ref{eq2aa}) again do not admit analytical solutions for arbitrary $\omega$. However, for $\omega =0$, there is
the solution%
\begin{equation}
\mu_0 \left( x\right) =\sin \left[ 2\arctan %
\left[ \exp \left( \eta \left( y_{0}+\left( x-x_{0}\right) -\frac{\lambda }{%
\alpha \left( 1+\lambda \right) }\tanh \left[ \alpha \left( x-x_{0}\right) %
\right] \right) \right) \right] \right]\text{,}
\end{equation}%
with $\psi_{0}=0$.

\textit{3. The case with trivial }$\phi $. We now consider that $\phi$ is in its fundamental configuration. In this case, whether we choose $%
\phi =0$ or $\phi =\pi $, the BPS Eq. (\ref{bpsphi}) is automatically satisfied. On the other hand, Eq. (\ref{bpschi}) gives
the well-known result%
\begin{equation}
\chi _{k}\left( x\right) =\tanh \left[ \alpha \left( x-x_{0}\right) \right] 
\text{.}
\end{equation}%

The linearized Eqs. (\ref{le1a}) and (\ref{le2a}) decouple. The two individual equations are%
\begin{equation}
\frac{d^{2}\mu }{dx^{2}}-\frac{2\alpha \lambda \chi _{k}\left( 1-\chi
_{k}^{2}\right) }{1+\lambda \chi _{k}^{2}}\frac{d\mu }{dx}+\left[ \omega
^{2}-\frac{\eta ^{2}}{\left( 1+\lambda \right) ^{2}}\left( 1+\lambda \chi
_{k}^{2}\right) ^{2}\right] \mu =0\text{,}  \label{deq1a}
\end{equation}%
\begin{equation}
\frac{d^{2}\psi }{dx^{2}}+\left[ \omega ^{2}+2\alpha ^{2}\left( 1-3\chi
_{k}^{2}\right) \right] \psi =0\text{.}  \label{deq2a}
\end{equation}%
Notably, they present the same structure of Eqs. (3.8) and (3.9) of Ref. \cite{joao}. In particular, when $\eta=2$, the Eqs. (\ref{deq1a}) and (\ref{deq2a}) recover exactly those expressions.

Equation (\ref{deq2a}) predicts the existence of one translational mode and one vibrational mode (with $\omega ^{2}=3\alpha ^{2}$). These solutions are%
\begin{equation}
\psi _{0}(x)=\text{sech}^{2}\left[ \alpha \left( x-x_{0}\right) \right]
\text{,}
\end{equation}%
\begin{equation}
\psi _{1}(x)=\tanh \left[ \alpha \left( x-x_{0}\right) \right] \text{sech}%
\left[ \alpha \left( x-x_{0}\right) \right] \text{,}
\end{equation}%
respectively.


\begin{figure*}[!ht]
\begin{center}
  \centering
    \subfigure[]{\includegraphics[width=0.42 \textwidth]{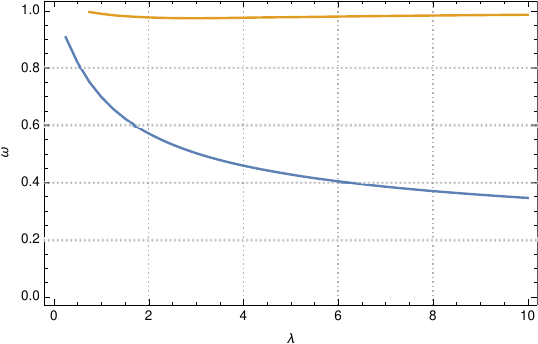}\label{fig5d0}}
    \subfigure[]{\includegraphics[width=0.42 \textwidth]{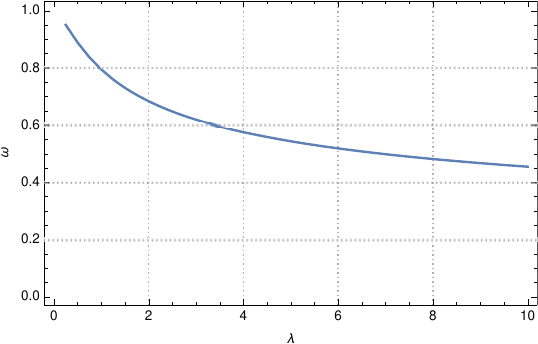}\label{fig5e0}}
    \subfigure[]{\includegraphics[width=0.42 \textwidth]{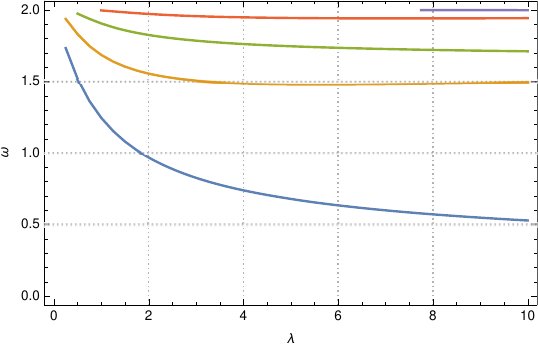}\label{fig5g0}}
    \subfigure[]{\includegraphics[width=0.42 \textwidth]{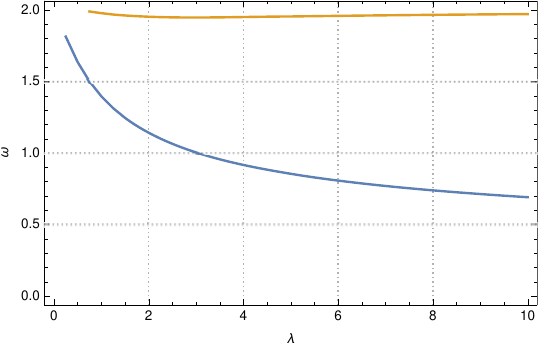}\label{fig5h0}}
  \caption{Discrete frequencies $\omega$ as functions of $\lambda$ for $\alpha=0.5$ (first column) and $\alpha=1.0$ (second column). Here, we have used $\eta=1.0$ (first row) and $\eta=2.0$ (second row). A nonvanishing $\lambda$ supports at least one vibrational mode.}
  \label{fig50}
\end{center}
\end{figure*}

Moreover, depending on $\eta$, $\alpha$ and $\lambda$, Eq. (\ref{deq1a}) may lead to additional vibrational modes. We have solved it numerically. Some of the results are plotted in Fig. (\ref{fig50}). This Figure shows the discrete frequencies $\omega$ as functions of $\lambda$ for different $\alpha$ and $\eta$. In general, a nonzero $\lambda$ always supports at least one vibrational mode. In addition, when $\eta=2.0$, the number of modes increases with $\lambda$. On the other hand, additional simulations have indicated that this number remains constant when $\eta=0.5$, for all $\alpha$. Note that, for $\alpha$ fixed, the number of modes increases with $\eta$.

\vspace{.55cm}

{\subsection{Collisions}}

We now investigate the scattering between the BPS kinks. First, we note that Eqs. (\ref{emplr1}) and (\ref{emclr1}) can be written as%
\begin{equation}
\frac{\partial ^{2}\phi }{%
\partial t^{2}}-\frac{\partial ^{2}\phi }{\partial x^{2}} -\frac{%
2\lambda \chi  }{1+\lambda \chi ^{2}
}\left( \frac{\partial \chi }{\partial t}\frac{\partial \phi }{\partial t%
}-\frac{\partial \chi }{\partial x}\frac{\partial \phi }{\partial x}\right) +%
\frac{\eta ^{2}\left( 1+\lambda \chi ^{2}\right)^2 }{2(1+\lambda)^2 } \sin \left[
2\phi \right]  =0\text{,}\label{eqmotion1}
\end{equation}%
\begin{equation}
\frac{\partial ^{2}\chi }{\partial t^{2}}-\frac{\partial ^{2}\chi }{\partial
x^{2}}-\frac{\lambda \chi \left( 1+\lambda \right) }{\left( 1+\lambda \chi
^{2}\right) ^{2}}\left[ \left( \frac{\partial \phi }{\partial x}\right)
^{2}-\left( \frac{\partial \phi }{\partial t}\right) ^{2}\right] +\frac{\eta
^{2}\lambda \chi }{1+\lambda }\sin ^{2}\left[ \phi \right] -2\alpha ^{2}\chi
\left( 1-\chi ^{2}\right) =0\text{,}\label{eqmotion2}
\end{equation}%
respectively. Here, we have used Eqs. (\ref{fx100a1}) and (\ref{p00}).

We study the collision of a kink-antikink pair symmetrically positioned around the origin. Therefore, we choose the initial conditions as
\begin{eqnarray}
\chi\left(x,0,x_0,v\right) &=& \chi_{K}\left( x+ x_{0},0,v\right)  + \chi_{\bar{K}}\left( x-x_0,0,-v\right) -1 \text{,}\label{eqaa}\\
\dot{\chi}\left(x,0,x_0,v\right) &=& \dot{\chi}_{K}\left( x+ x_{0},0,v \right) + \dot{\chi}_{\bar{K}} \left(x-x_0,0,-v \right)\text{,}
\end{eqnarray}%
and
\begin{eqnarray}
\phi \left(x,0,x_0,v\right) &=&\phi_{K}\left( x+ x_{0},0,v \right) +\phi_{\bar{K}} \left(x- x_{0}, 0,-v\right) -\pi\text{,}\label{eqbb}\\
\dot{\phi} \left(x,0,x_0,v\right) &=&\dot{\phi}_{K}\left( x+ x_{0},0,v \right) +\dot{\phi}_{\bar{K}} \left(x-x_{0},0,-v\right)\text{.}
\end{eqnarray}%
These expressions mean that the BPS kinks, given by Eqs. (\ref{schik001a1}) and (\ref{sphik001a1}), and antikinks,
given by Eqs. (\ref{schiak001a1}) and (\ref{sphiak001a1}), now move against each other with velocity $v>0$. Here, $\phi_K(x,t,v)=\phi_K(\gamma(x-vt))$ and $\chi_K(x,t,v)=\chi_K(\gamma(x-vt))$, where $\gamma=(1-v^2)^{-1/2}$ is the Lorentz factor. The constant terms in the r.h.s of Eqs. (\ref{eqaa}) and (\ref{eqbb}) were introduced
so that the initial configuration is in the same topological BPS sector, see Fig. \ref{times_A_sigma_1000}.

In what follows, we describe the scattering process. We solve Eqs. (\ref{eqmotion1}) and (\ref{eqmotion2}) in a box $-z_{max}<x<z_{max}$. We consider $z_{max}=200$ and a space step $\delta x=0.05$. Partial derivatives with respect to $x$ were approximated using the five-point stencil. The resulting equations were integrated via the fifth-order Runge-Kutta method with adaptive step size. We have considered periodic boundary conditions and $x_0=10$ fixed.


We study collision for different values of $\lambda$ and $\alpha$ in Eqs. (\ref{eqmotion1}) and (\ref{eqmotion2}), with $\eta=1$ fixed. Note that, due to the coupling between $\phi$ and $\chi$, the scattering occurs simultaneously in both sectors.

\begin{figure*}[!ht]
\begin{center}
  \centering
    \subfigure[]{\includegraphics[width=0.48 \textwidth]{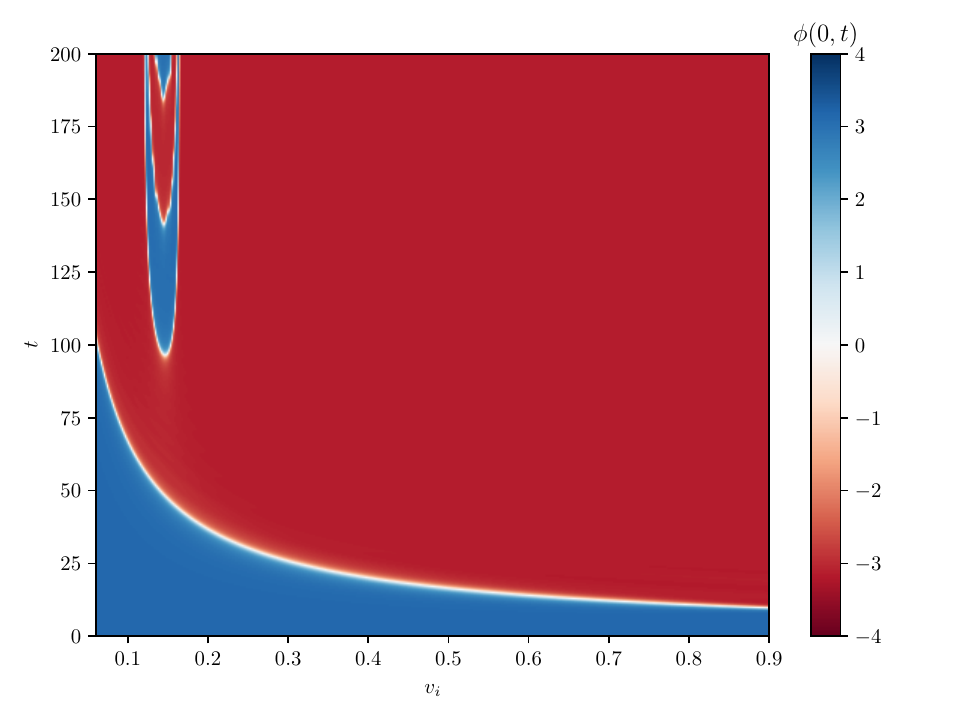}\label{fig5a}}
    \subfigure[]{\includegraphics[width=0.48 \textwidth]{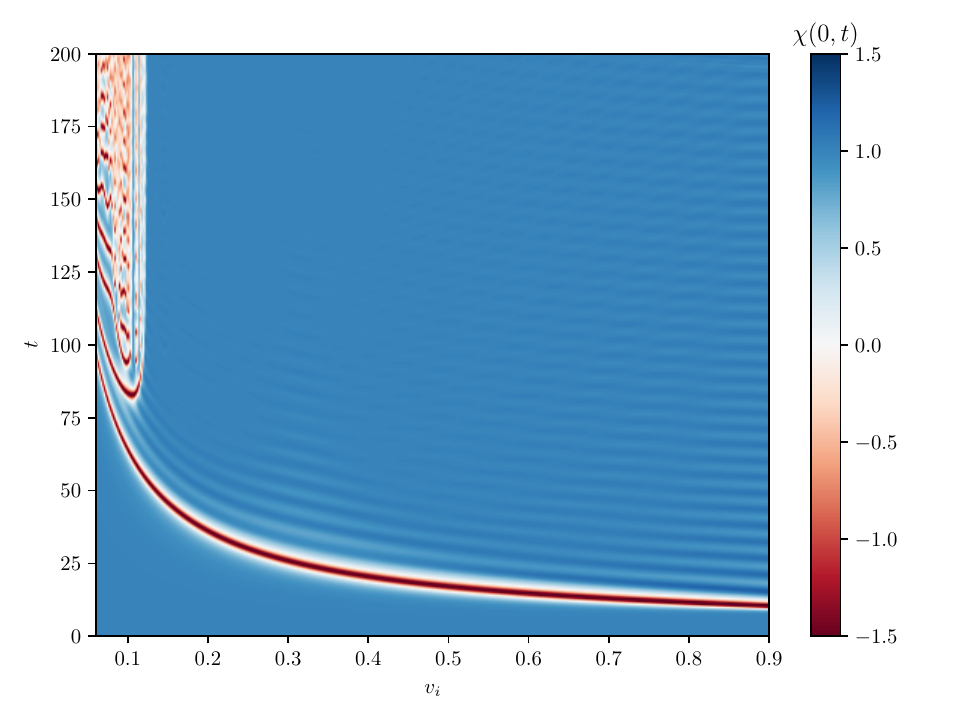}\label{fig5b}}
    \subfigure[]{\includegraphics[width=0.48 \textwidth]{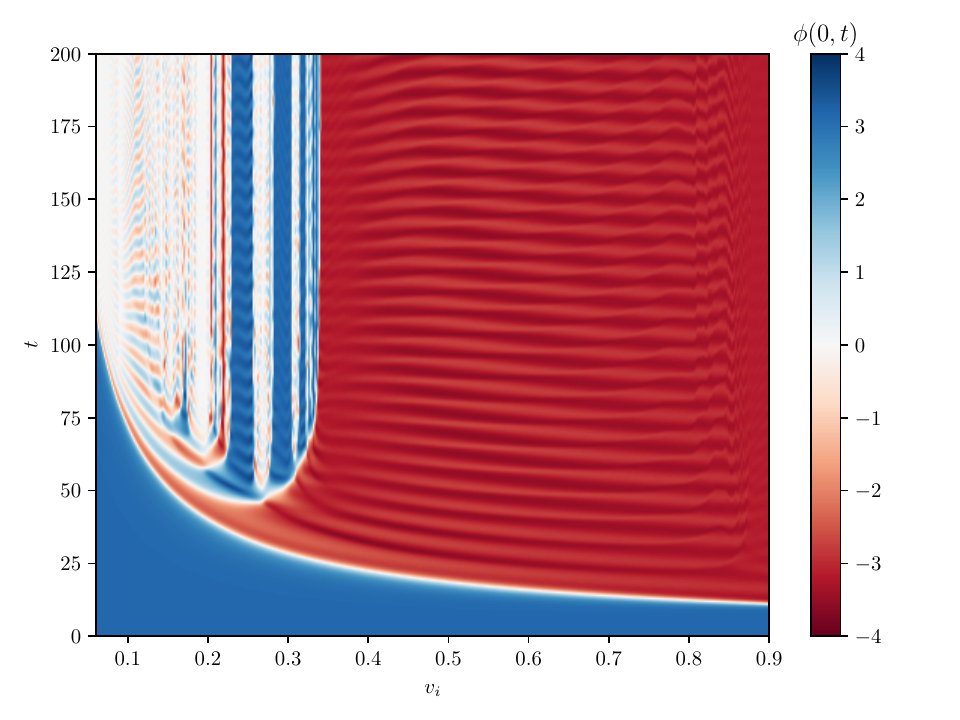}\label{fig5g}}
    \subfigure[]{\includegraphics[width=0.48 \textwidth]{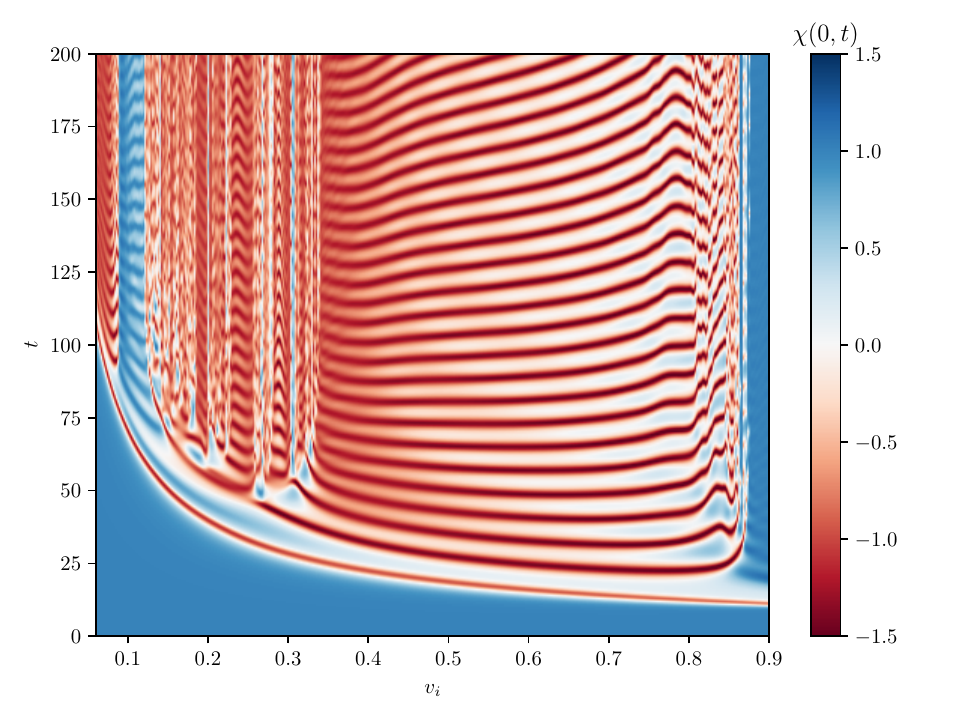}\label{fig5h}}
  \caption{Field values at the center of mass as functions of $t$ and $v$ for $\lambda=0.01$ (first row) and $\lambda=0.5$ (second row). We have fixed $\alpha=0.5$.}
  \label{fig5}
\end{center}
\end{figure*}

Initially, we have investigated the scattering for different $v$ and $\lambda$. In this case, we have fixed $\alpha=0.5$. The resonant structures appear in Fig. \ref{fig5} where the field values at the center of mass are shown. The vertical axis represents time, while the horizontal one stands for the velocity of the colliding structures.

Small $\lambda$ means a weak coupling, i.e. a small contribution from geometric constrictions. See Fig. \ref{fig5b}, for instance. It presents a thin resonance window for $\lambda=0.01$. It is important to clarify the colormap in this Figure. The horizontal red line denotes a collision (i.e. a bounce), while the vertical blue stripes represent a kink-antikink pair.

As $v$ increases and approximates a critical value, the second horizontal red line diverges. From this point on, only one bounce is observed. We recognize some similarities between this structure and that of the $\phi^4$ model, as expected for small $\lambda$.

In Fig.~\ref{fig5a}, the scattering of $\phi$ exhibits a novel aspect. In this example, for both small and high $v$, the kink-antikink pair approaches, collides, and then splits in the opposite topological sector. This is the usual sine-Gordon behavior. However, within a very small range of $v$, a new structure emerges. It indicates a continuous interpolation between the two topological sectors. This behavior represents a novelty. In the absence of constrictions, the standard sine-Gordon model does not support such a behavior.

\begin{figure*}[!ht]
\begin{center}
  \centering
    \subfigure[]{\includegraphics[width=1.0 \textwidth]{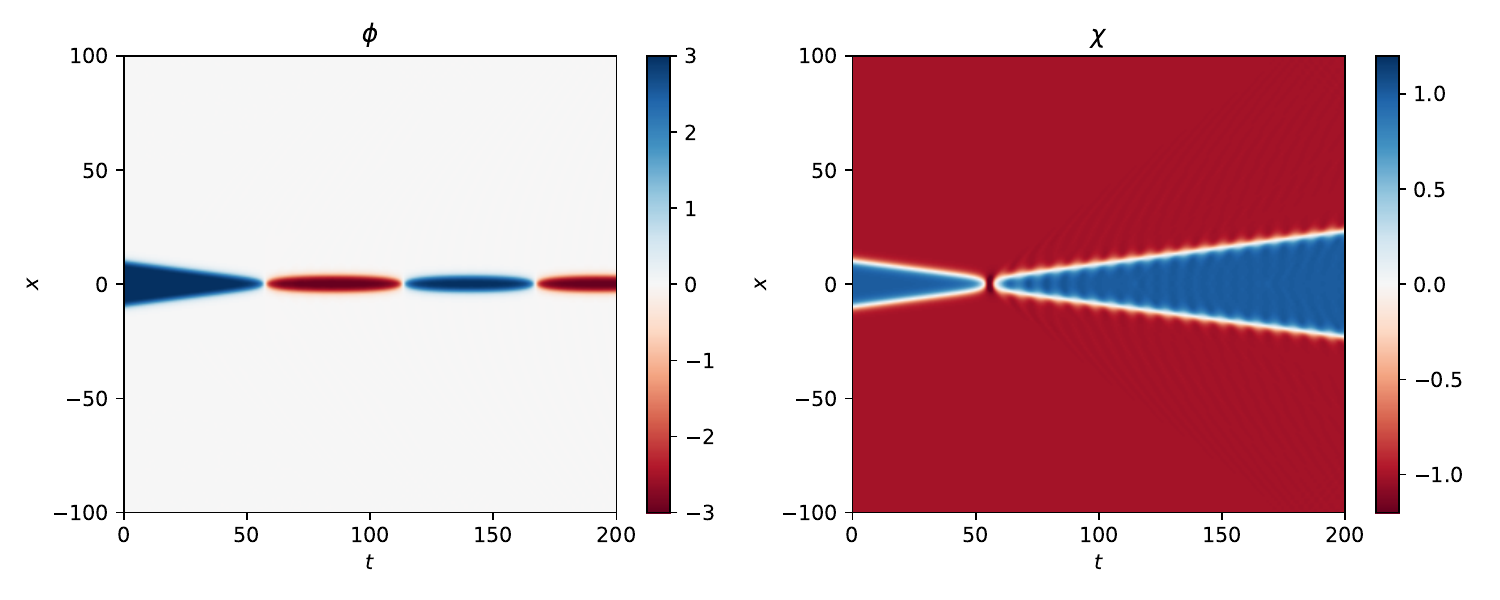}\label{col1A}}
    \subfigure[]{\includegraphics[width=1.0 \textwidth]{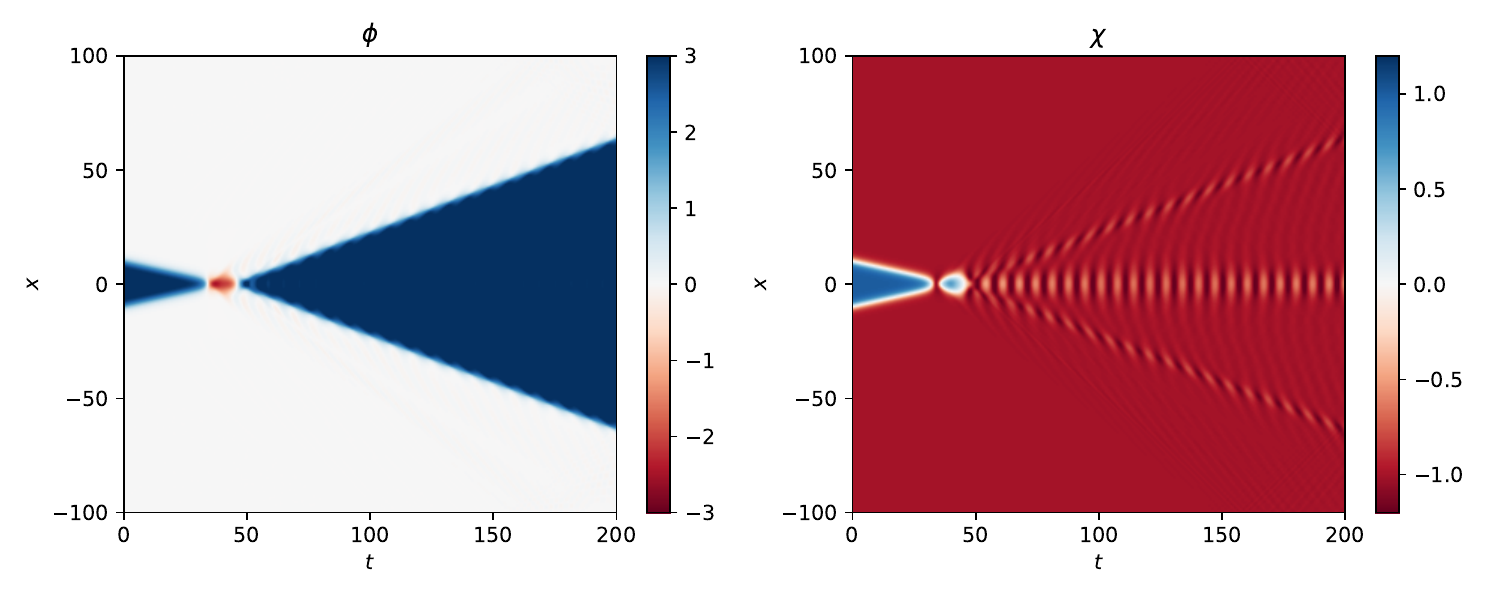}\label{col1B}}
  \caption{Field evolution in spacetime for $\lambda=0.01$ and $v_i=0.14$ (first row), and $\lambda=0.5$ and $v_i=0.24$ (second row). We have fixed $\alpha=0.5$.}
  \label{col1}
\end{center}
\end{figure*}

The interaction between $\chi$ and $\phi$ is responsible for this new output. The behavior can be seen in Fig. \ref{col1A}. We have chosen $\lambda=0.01$ and $v_i=0.14$. The kink-antikink pair in $\phi$ collides and then initiates a continuous interpolation between the topological sectors. In contrast, $\chi$ performs an inelastic scattering followed by a small emission of radiation.

As $\lambda$ increases, the coupling gets stronger. As a consequence, the outcome of a kink-antikink scattering is significantly altered. Then, interesting new patterns appear. For instance, when $\lambda=0.5$, Fig.~\ref{fig5g} shows the formation of a novel resonant structure for $\phi$. At the same time, Fig.~\ref{fig5h} indicates the almost total annihilation of $\chi$.

In Fig.~\ref{fig5g}, vertical blue stripes appear for $0.2<v_i<0.35$. They correspond to the emergence of a kink-antikink pair. This behaviour is illustrated in Fig.~\ref{col1B}. In this case, the pair approaches and collides a first time. Then, it changes phase and collides again. Finally, a kink-antikink pair scatters in the initial topological sector. 


In addition, Fig.~\ref{fig5h} reveals a suppression of the resonant window. The result is a large area that corresponds to the appearance of either oscillating pulses or a bion scattering. Fig.~\ref{col1B} illustrates this behavior.

\begin{figure*}[!ht]
\begin{center}
  \centering
    \subfigure[]{\includegraphics[width=0.48 \textwidth]{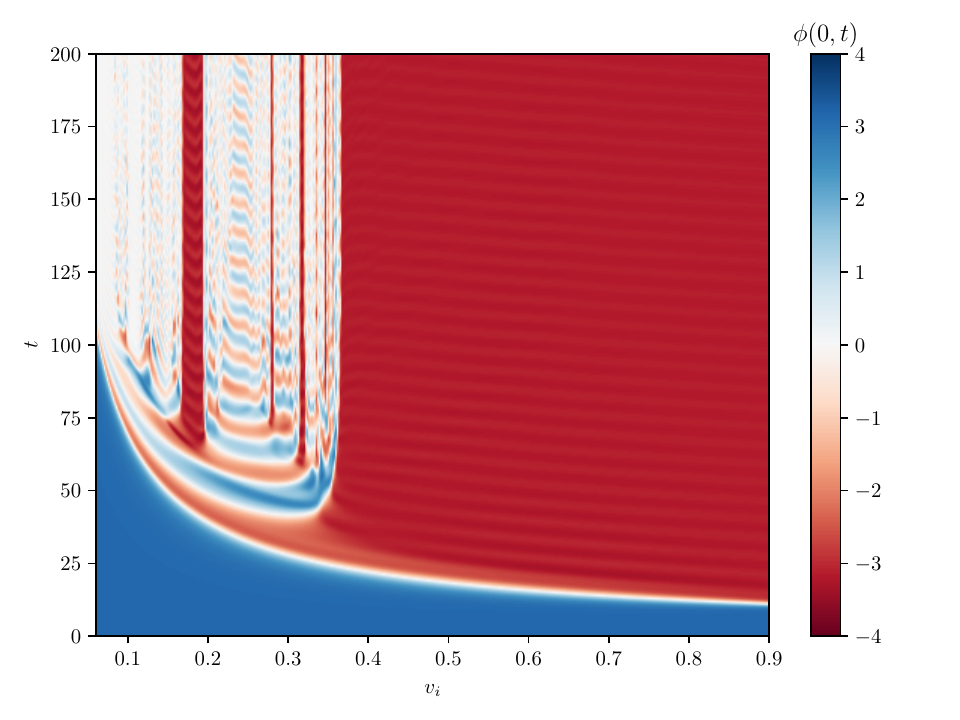}\label{fig51A}}
    \subfigure[]{\includegraphics[width=0.48 \textwidth]{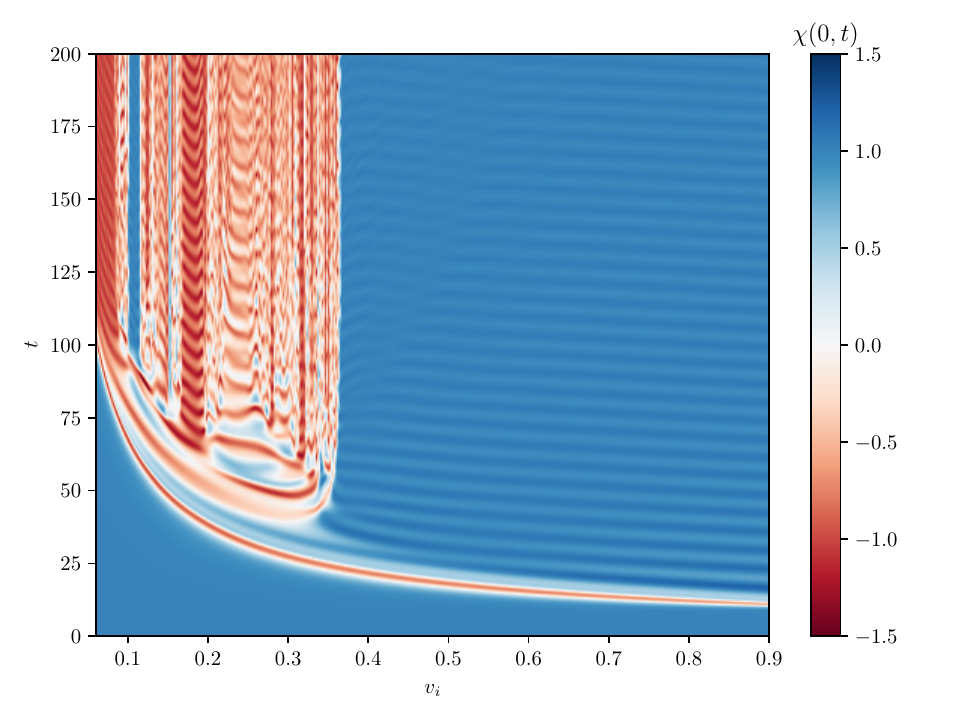}\label{fig51B}}
    \subfigure[]{\includegraphics[width=0.48 \textwidth]{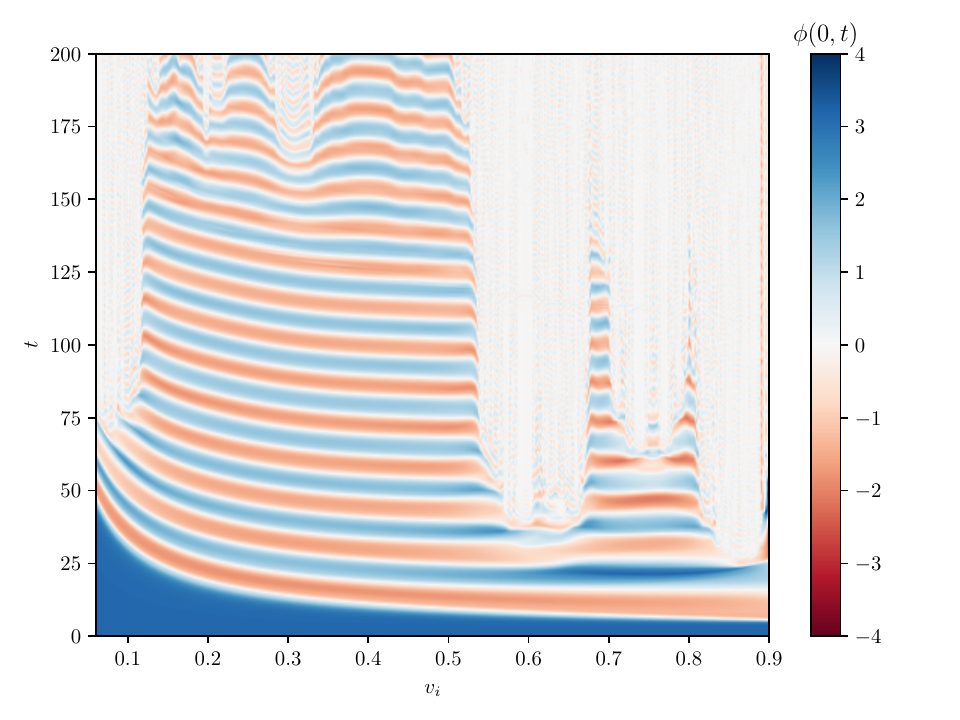}\label{fig51G}}
    \subfigure[]{\includegraphics[width=0.48 \textwidth]{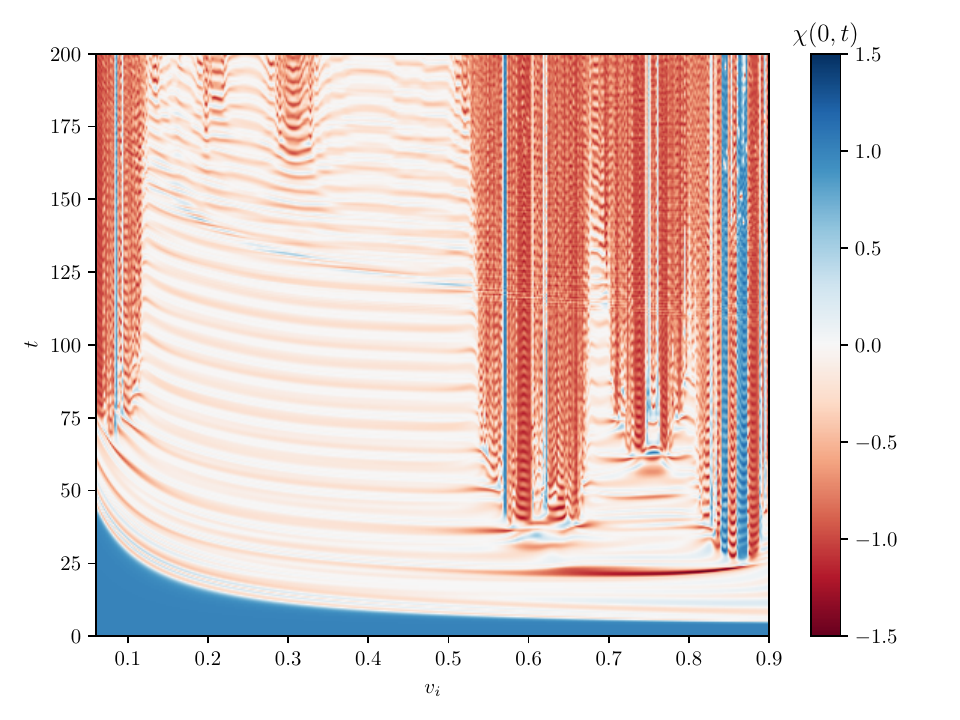}\label{fig51H}}
  \caption{Field values at the center of mass as functions of $t$ and $v$ for $\lambda=1.0$ (first row) and $\lambda=4.0$ (second row). We have fixed $\alpha=0.5$.}
  \label{fig51}
\end{center}
\end{figure*}


Obviously, as $\lambda$ increases, it continuously modifies the resonant structures. See Fig.~\ref{fig51}, for example. It depicts the results for $\lambda=1$ and $\lambda=4$. We still notice some resonance window for $\phi$.

In particular, Fig.~\ref{fig51A} reveals a new behavior. For instance, when $v_i=0.19$, the sine-Gordon field collides and then begins to interpolate between the topological sectors (blue and red colors). After this interpolation, it finally scatters out in a different vacuum state (vertical red band).

\begin{figure*}[!ht]
\begin{center}
  \centering
    \subfigure[]{\includegraphics[width=1.0 \textwidth]{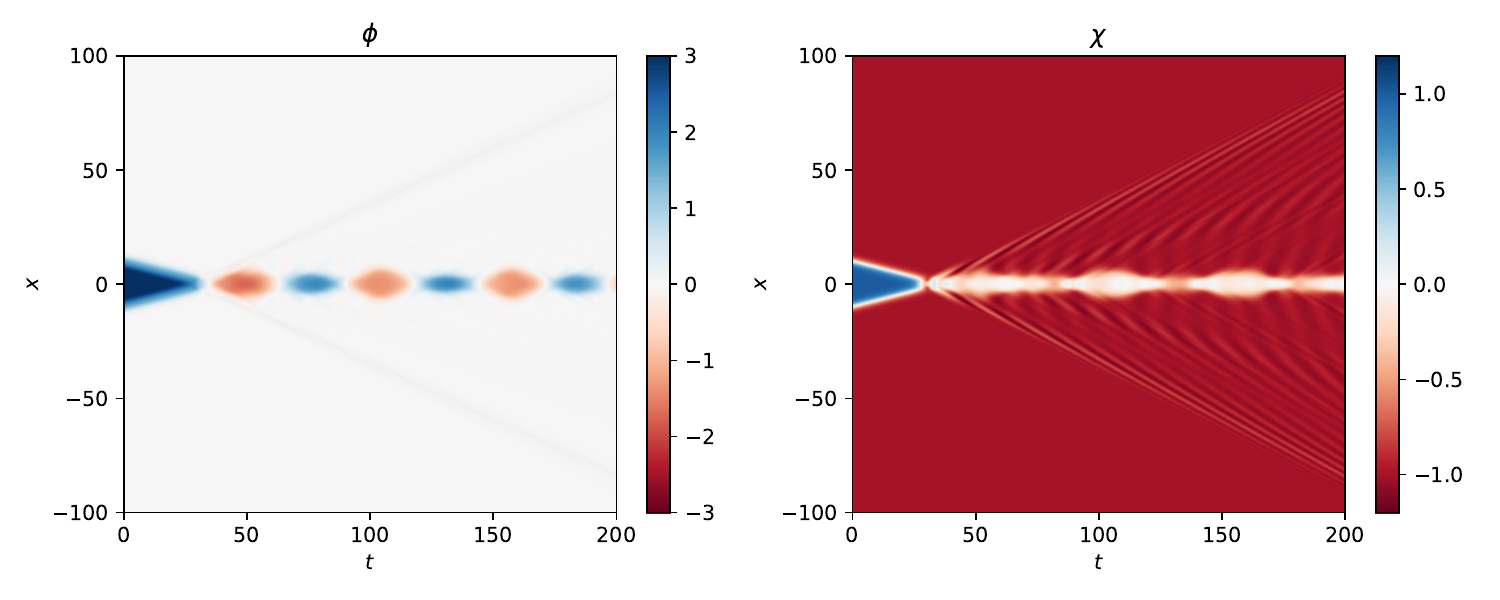}\label{fig52a}}
    \subfigure[]{\includegraphics[width=1.0 \textwidth]{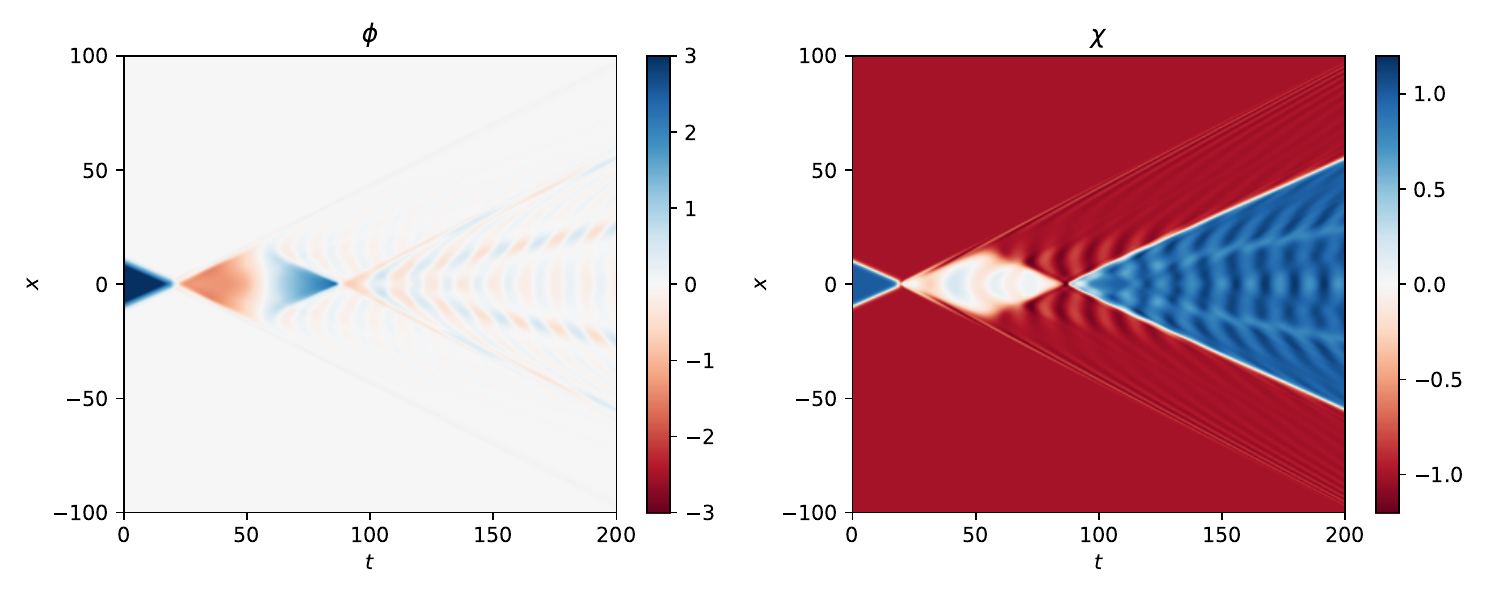}\label{fig52b}}
  \caption{Field evolution in spacetime for $v_i=0.5$ (first row) and $v_i=0.865$ (second row). We have fixed $\lambda=4.0$ and $\alpha=0.5$.}
  \label{fig52}
\end{center}
\end{figure*}

As $v$ increases, a wide one-bounce region is formed. The same applies for $\chi$, see Fig.~\ref{fig51B}. In this case, however, the pair separates in the same initial vacuum after a single collision. In addition, there is a region that indicates the annihilation of the pair. A very narrow blue band that means the kink-antikink separation also appears.

The scattering structure for $\lambda=4.0$ is presented in Figs.~\ref{fig51G} and \ref{fig51H}. They demonstrate how an increasing $\lambda$ significantly alters the system. Interestingly, there are no one-bounce windows. Therefore, the critical velocity cannot be determined. Furthermore, only thin higher-bounce windows are displayed for $\chi$. In this scenario, the kink-antikink collision leads to annihilation, regardless how massive the kinks are.

\begin{figure*}[!ht]
\begin{center}
  \centering
    \subfigure[]{\includegraphics[width=0.48 \textwidth]{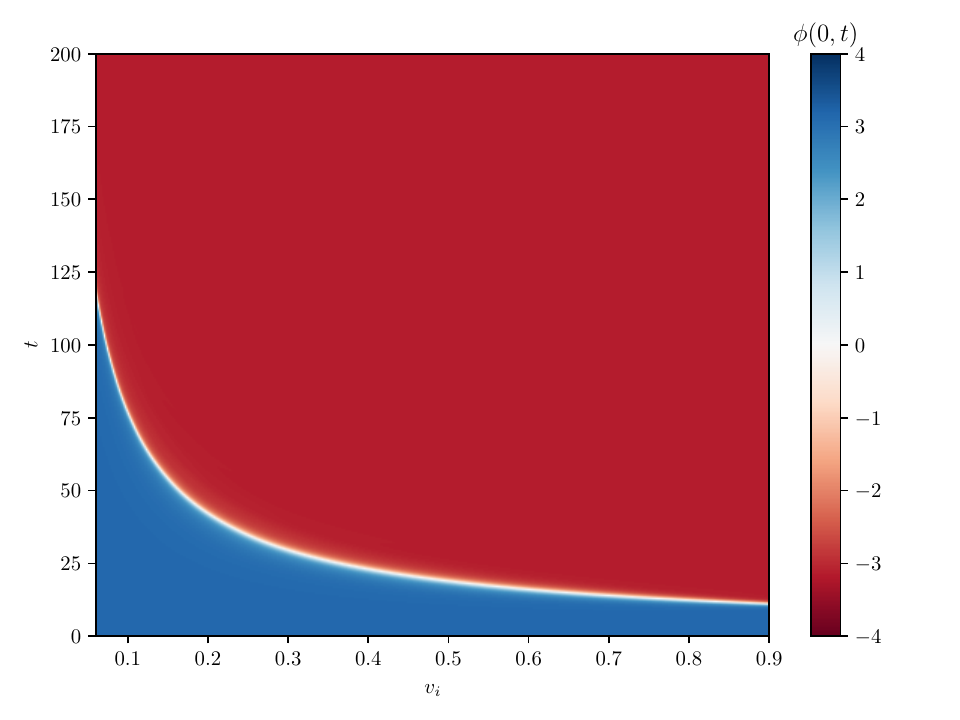}\label{fig6a}}
    \subfigure[]{\includegraphics[width=0.48 \textwidth]{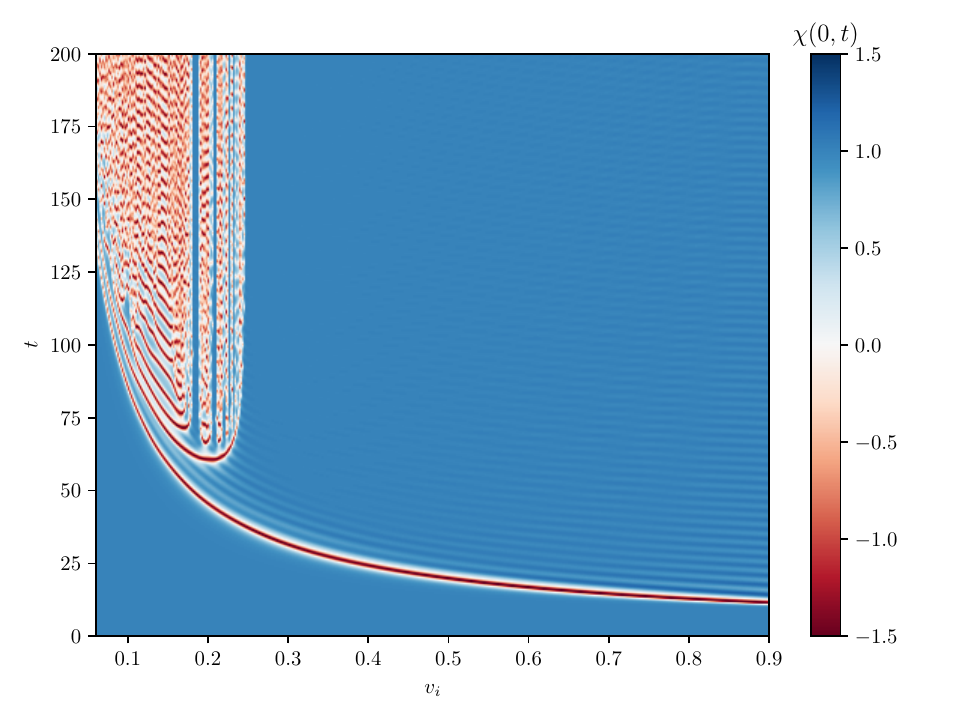}\label{fig6b}}
    \subfigure[]{\includegraphics[width=0.48 \textwidth]{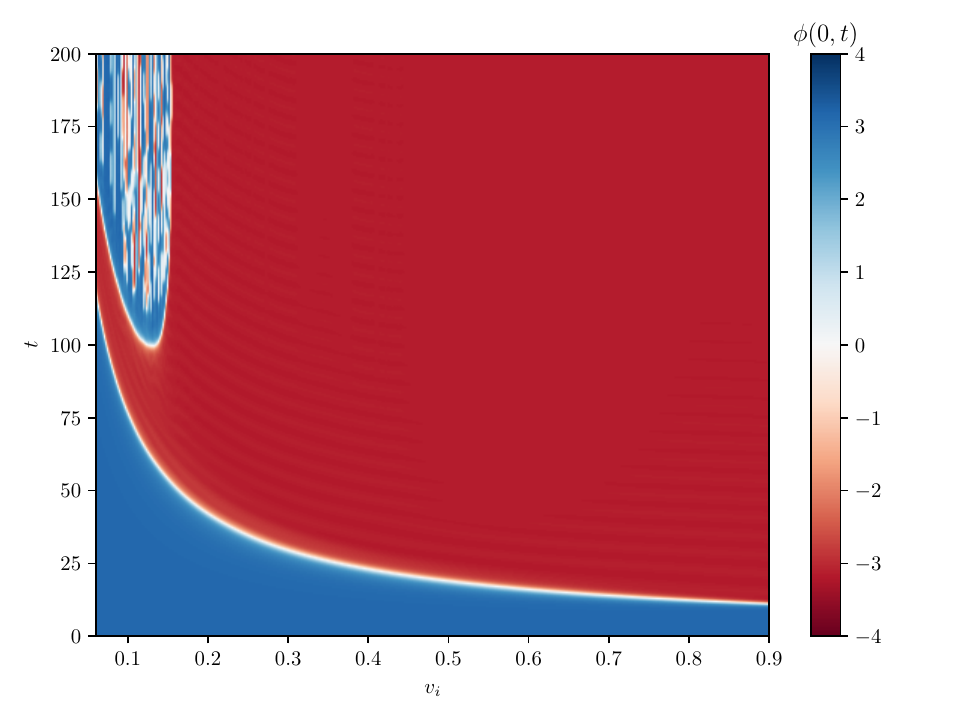}\label{fig6c}}
    \subfigure[]{\includegraphics[width=0.48 \textwidth]{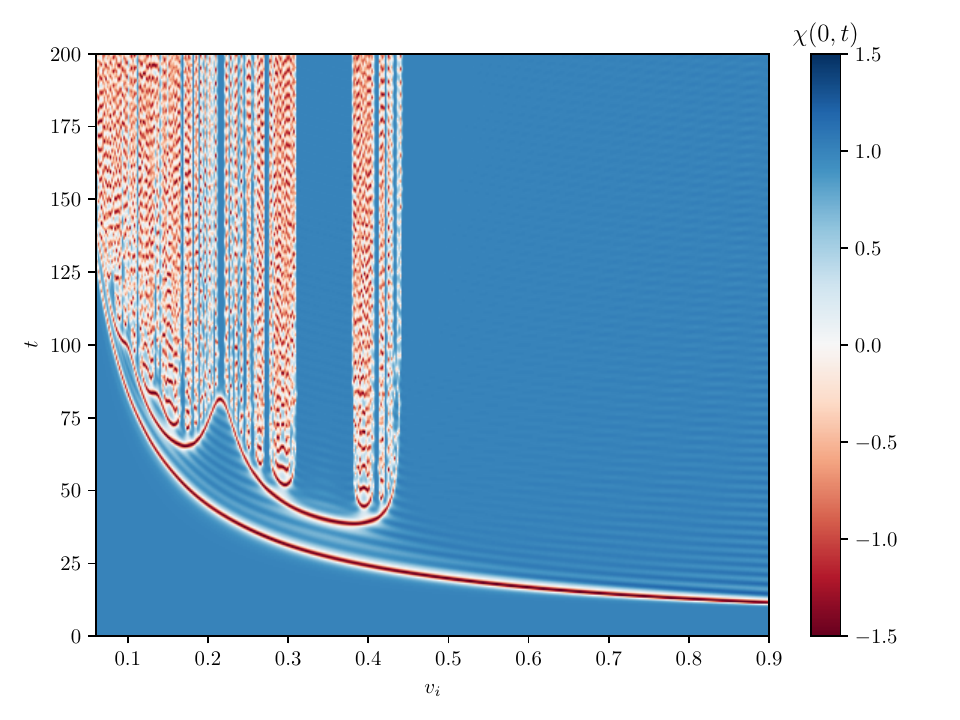}\label{fig6d}}
  \caption{Scalar fields at the center of mass as functions of $t$ and $v$ for $\lambda=0.01$ (first row) and $\lambda=0.05$ (second row). We have fixed $\alpha=1.0$.}
  \label{fig6}
\end{center}
\end{figure*}


In Fig.~\ref{fig52}, we plotted some examples of the evolution reported in Fig.~\ref{fig51}. Regarding this, Fig.~\ref{fig52a} shows a complete annihilation of the original pair. At higher $v$, Fig.~\ref{fig52b} reveals that $\phi$ and $\chi$ decay into oscillating pulses and a kink-antikink pair, respectively. In the latter case, the kinks almost annihilate themselves before the emergence of the final pair.

We also consider the effects related to variations on $\alpha$. In particular, Fig.~\ref{fig6} depicts the resonant structure for $\alpha=1.0$. Now, the mass of $\chi$ is greater than before. Again, $\lambda$ modulates the impact due to geometric constrictions. At higher $v$, the pairs collide once and then scatter. In Fig.~\ref{fig6a}, we only notice a phase shift after that single collision. Here, $\lambda=0.01$ is small. The result then mimics the standard sine-Gordon behavior. On the other hand, Fig.~\ref{fig6b} reveals resonance windows and a critical velocity. This behavior is comparable to a kink-antikink scattering in the canonical $\phi^4$ model.

Obviously, the system gets more intricate as $\lambda$ increases. The corresponding modifications are shown in Figs.~\ref{fig6c} and \ref{fig6d}. As the coupling gets stronger, a kink-antikink pair emerges in $\phi$ after a few collisions for small $v$, see Fig.~\ref{fig6c}.

\begin{figure*}[!ht]
\begin{center}
  \centering
    \subfigure[]{\includegraphics[width=0.48 \textwidth]{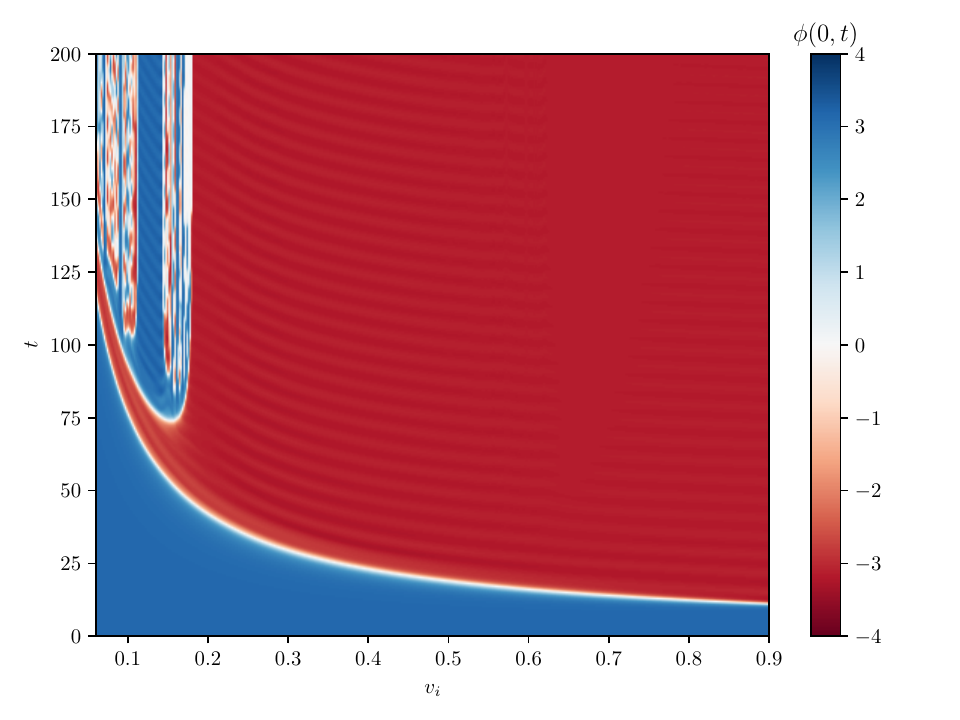}\label{fig62a}}
    \subfigure[]{\includegraphics[width=0.48 \textwidth]{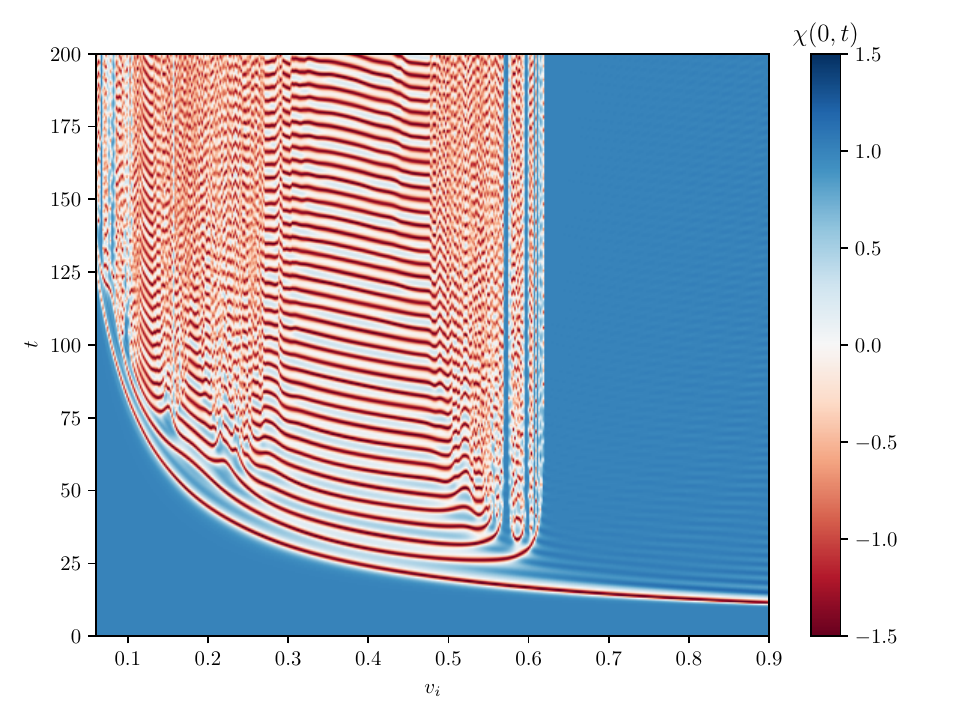}\label{fig62b}}
    \subfigure[]{\includegraphics[width=0.48 \textwidth]{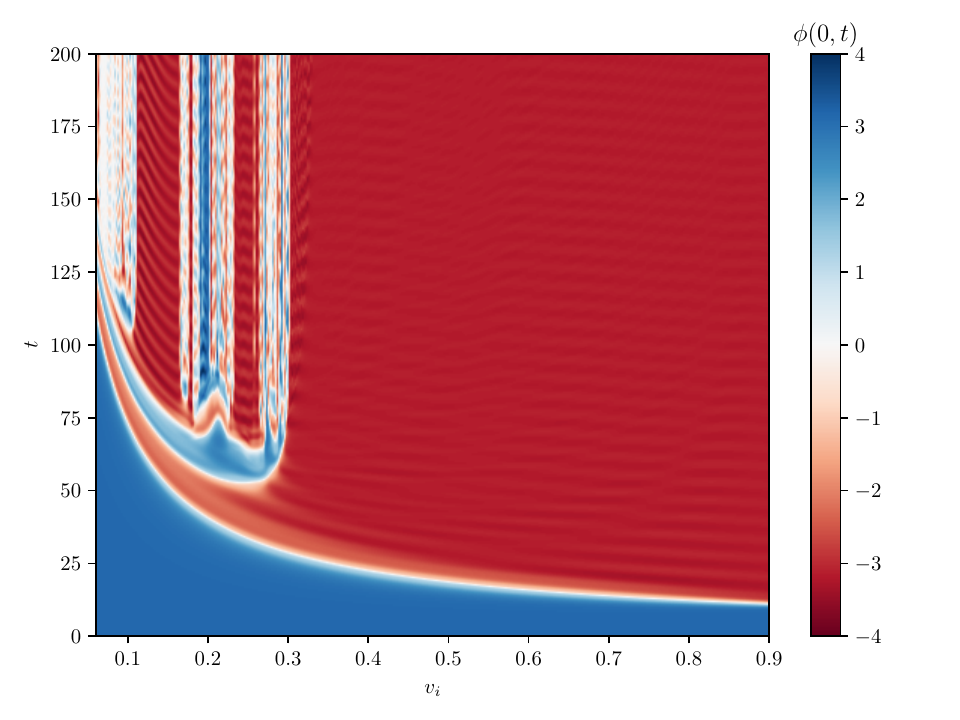}\label{fig62c}}
    \subfigure[]{\includegraphics[width=0.48 \textwidth]{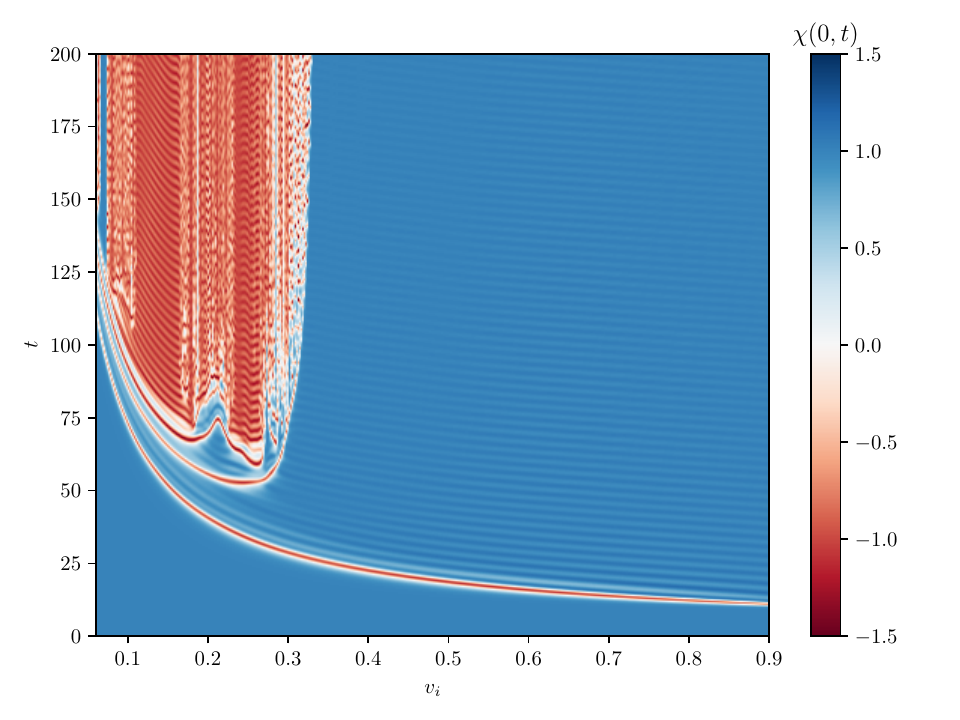}\label{fig62d}}
  \caption{Scalar fields at the center of mass as functions of $t$ and $v$ for $\lambda=0.1$ (first row) and $\lambda=2.0$ (second row). We have fixed $\alpha=1.0$.}
  \label{fig62}
\end{center}
\end{figure*}


In contrast to Fig.~\ref{fig6b}, the resonance structure for $\chi$ is entirely different, see Fig.~\ref{fig6d}. Now, a deformed two-bounce window appears around $v_i \approx 0.22$. There are also narrow resonant windows around a large vertical blue band for $0.3<v_i<0.4$.



As $\lambda$ increases, it favors the occurrence of resonance windows for $\phi$. Simultaneously, it leads to the annihilation of the original kinks in $\chi$. For instance, when comparing Figs.~\ref{fig6d} and \ref{fig62b}, we see that the region which exhibits a bion behavior increases, while the two-bounce window is almost entirely suppressed.

\begin{figure*}[!ht]
\begin{center}
  \centering
    \subfigure[]{\includegraphics[width=0.48 \textwidth]{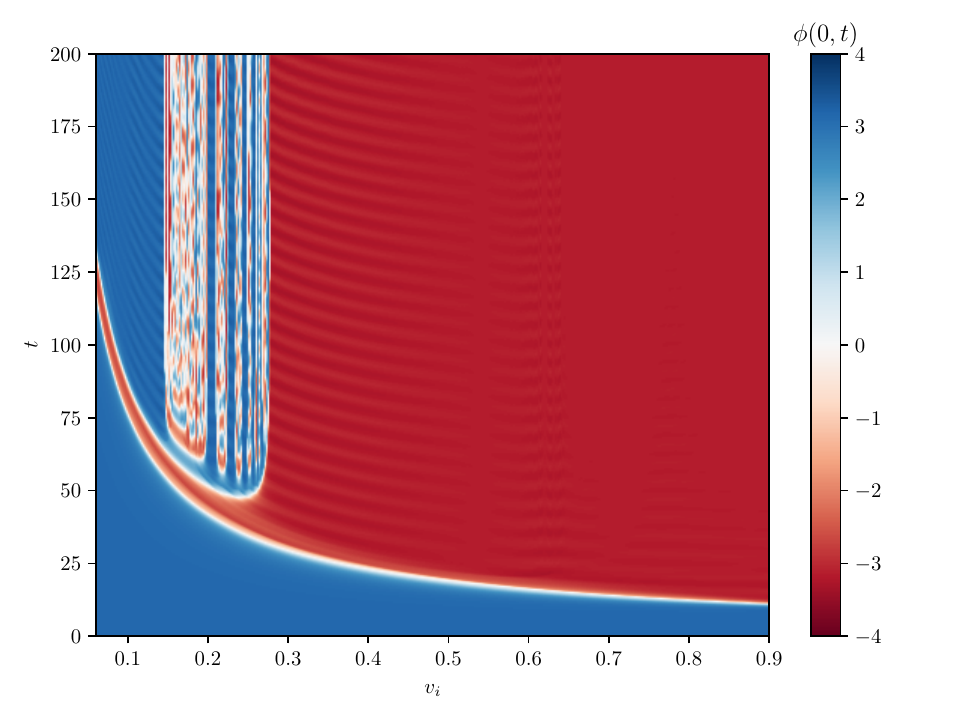}\label{fig7a}}
    \subfigure[]{\includegraphics[width=0.48 \textwidth]{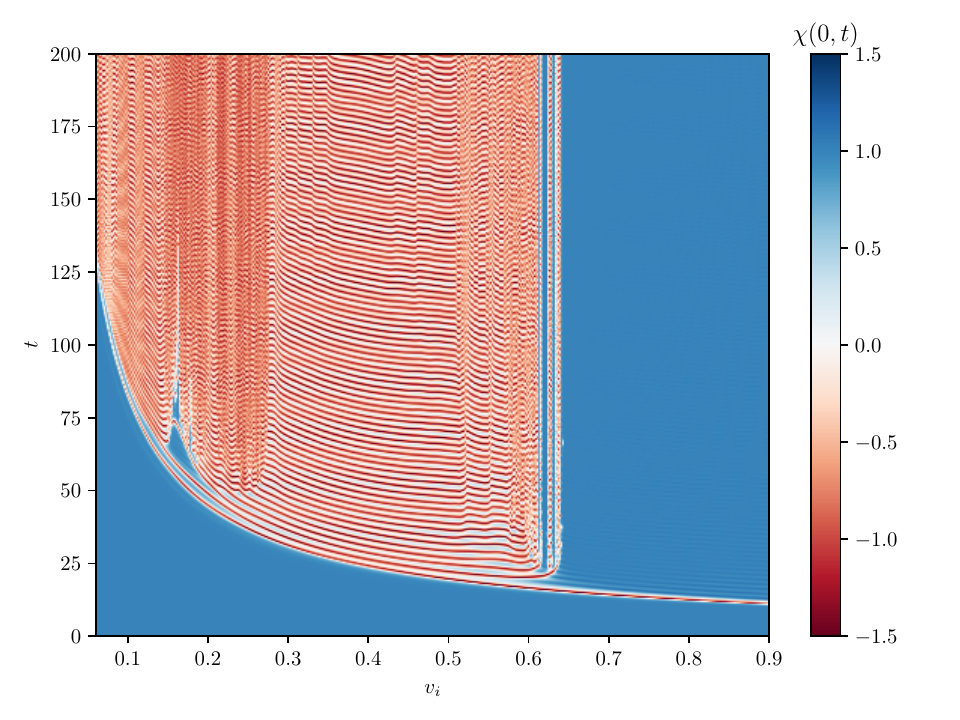}\label{fig7b}}
    \subfigure[]{\includegraphics[width=0.48 \textwidth]{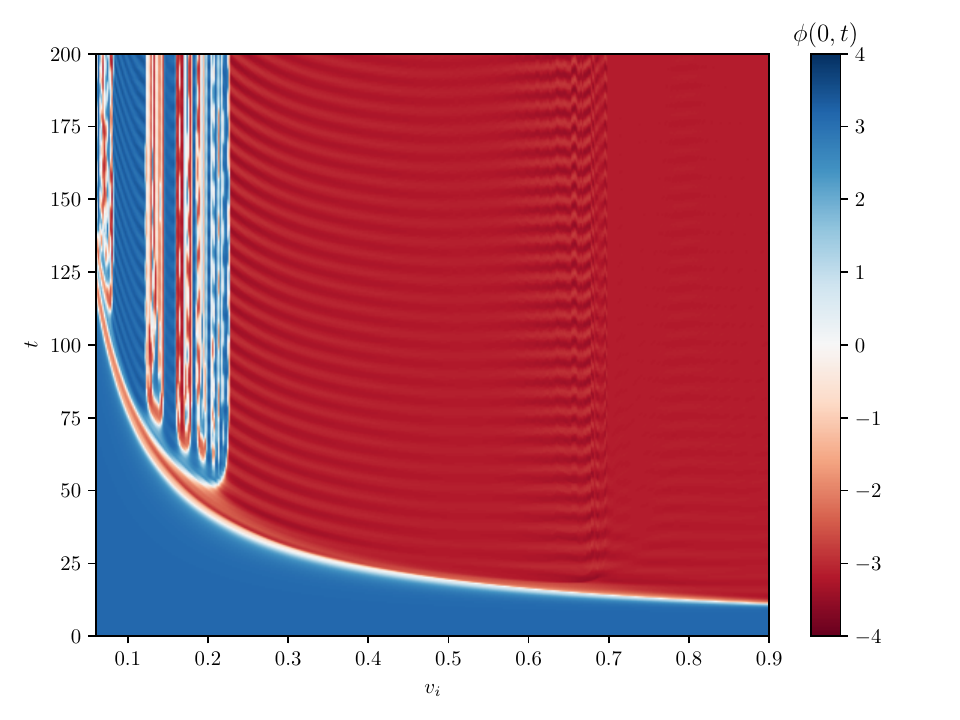}\label{fig7c}}
    \subfigure[]{\includegraphics[width=0.48 \textwidth]{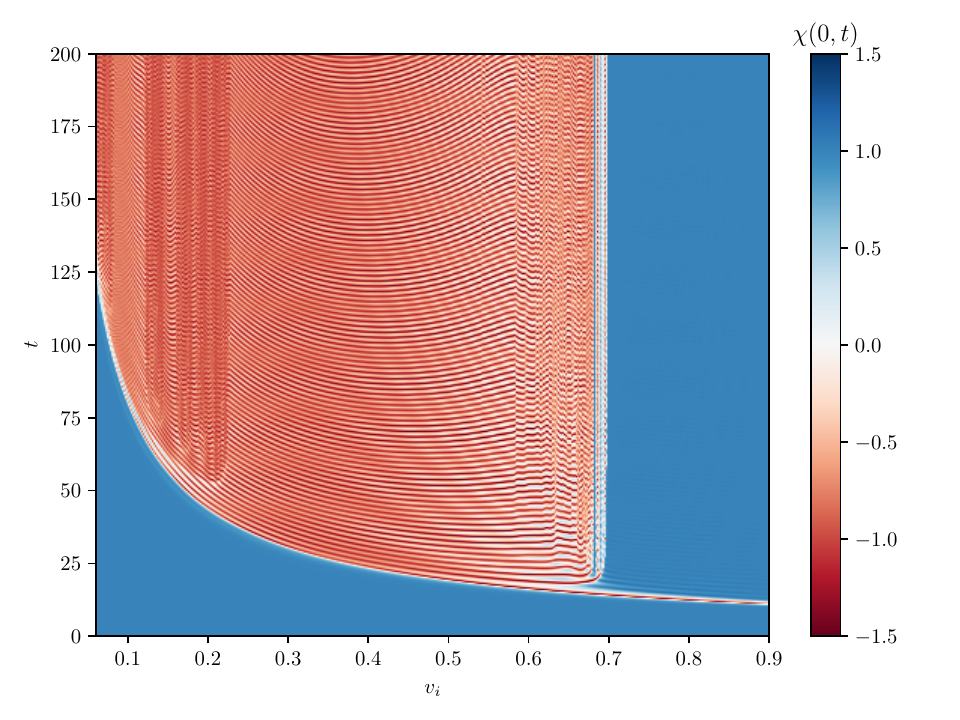}\label{fig7d}}
  \caption{Scalar fields at the center of mass as functions of $t$ and $v$ for $\lambda=0.5$ (first row) and $\lambda=1.0$ (second row). We have fixed $\alpha=2.0$.}
  \label{fig7}
\end{center}
\end{figure*}

Interestingly, an increasing $\lambda$ engenders the formation of new well-defined two-bounce windows for $\phi$. For instance, for $v_i=0.12$, Fig.~\ref{fig62a} presents a large blue band that indicates the separation of the pair after two collisions.

Figures \ref{fig62c} and \ref{fig62d} depict the scattering structure for $\lambda=2.0$. In this case, the constriction is even stronger. Therefore, it affects the structure much more. It is important to remember that now the $\phi$ field assumes a two-kink profile. This fact also contributes to the complexity of the results.

The critical velocity is now shared by both $\phi$ and $\chi$. For this last sector, we essentially identify annihilation and one-bounce. However, for $\phi$, there is an alternation between a bion behavior and resonant windows for small $v$.

In contrast to Fig.~\ref{fig62a}, an intriguing aspect may be observed in Fig.~\ref{fig62c}. Here, a few resonant windows mark the separation of the pair (vertical red band). In this region, the original pair collides and then it changes phase.


Finally, we briefly consider the results for $\alpha=2.0$. In this case, $\chi$ is more massive than $\phi$. Again, for small $\lambda$, our simulations reveal that the resonant structure remains essentially unchanged. In other words, there are only a phase shift in $\phi$ and two-bounce windows in $\chi$. However, as $\lambda$ increases, significant variations occur. 

Figure \ref{fig7} depicts the resonant structure for $\alpha=2$. The two-bounce windows are almost completely suppressed, while the bion area for $\chi$ is enlarged. On the other hand, the number of resonances for $\phi$ increases notably. 

\section{Summary and perspectives} \label{secIV}

We have investigated the effects on a sine-Gordon kink-like solution due to geometrical constrictions. We have clarified the mechanism through which these solutions develop internal structures. These structures mimic a two-kink profile. This result is a direct manifestation of the those constrictions.

We have considered an enlarged $(1+1)$-dimensional model with two real scalar fields, i.e. $\phi(x,t)$ and $\chi(x,t)$. These fields mutually interact via a nontrivial function $f(\chi)$. This function couples the field sectors and simulates the presence of geometrical constrictions.

We have minimized the total energy via the implementation of the BPS formalism. As a result, we have obtained the corresponding first-order equations and their BPS kink solutions.

In the sequence, we have studied how the shape of these profiles changes with a real parameter $\lambda$. This parameter enters the expression for $f(\chi)$, and therefore modulates the coupling between the fields. In general, for small $\lambda$ (weak interaction), the fields decouple and behave canonically. On the other hand, as $\lambda$ increases (strong coupling), the BPS sine-Gordon field develops a two-kink profile.

We have also explored the spectrum against small perturbations. Due to the high non-linearity of the system, we have concluded that it is quite hard to address the stability of the full scenario, even numerically.

In order to circumvent this issue, we have focused on a few particular cases. We have solved their linearized equations analytically. We have then concluded by the existence of both translational and vibrational modes. These modes play an important role during the scattering process via the energy exchange mechanism.

Finally, we have studied how the geometric constrictions affect a kink-antikink collision. Here, the scattering occurs simultaneously in both fields $\phi(x,t)$ and $\chi(x,t)$. We have fixed a constant mass for $\chi$. Regarding this, we have solved numerically the collision process for different $\lambda$. 

In general, for very small $\lambda$, we have observed that the kink-antikink pair scatters standardly. In other words, for a weak coupling between $\phi(x,t)$ and $\chi(x,t)$, the collisions evolve almost independently. However, even for small $\lambda$, we have identified interesting new results, as a bion behavior in the sine-Gordon sector $\phi$.

As $\lambda$ increases, on the other hand, the coupling between the fields gets stronger. In such a regime, the effects caused by geometrical constrictions are relevant. As a consequence of such influence, an intricate resonance structure and two-bounce windows have appeared for the sine-Gordon field $\phi$. We have also observed the occurrence of a large area that indicates a bion behavior for $\chi$.

In general, it is not clear that the coupling via an {\textit{arbitrary}} $f(\chi)$ works well for all possible choices of $W(\phi,\chi)$. On the other hand, the superpotential defines the expressions for $W_{\phi}$ and $W_{\chi}$. In particular, once one knows $W_{\chi}(\phi,\chi)$, Eq. (\ref{bpschi}) provides the BPS solution for $\chi(x)$.

At the same time, Eq. (\ref{bpsphi}) can always be rewritten in terms of $y$. As a consequence, once $W_{\phi}$ is known, it is possible to find a BPS solution $\phi(y)$. This result does not depend on $f$.

However, a physically acceptable solution $\phi(x)$ depends on Eq. (\ref{fc0a1}) itself. In other words, it depends on $f(\chi)$ and $\chi(x)$. Given these two expressions, that Equation is expected to provide a well-defined relation $y=y(x)$. Such a relation must allow us to rewrite $\phi(y)$ as a real expression $\phi(x)$.

For instance, suppose that now the sine-Gordon field is $\chi$, while $\phi$ is submitted to the $\phi^4$ potential. The first-order kinks are then $\chi _{k}\left( x\right) =2\arctan \left[ \exp \left( \alpha x \right) \right]$ and $\phi _{k}\left( y\right) =\tanh \left( \eta y \right)$. Whether these kinks are assumed to interact via $f(\chi)$ as given by Eq. (\ref{fx100a1}), then Eq. (\ref{fc0a1}) provides a complex solution $y(x)$. This solution does not allow us to rewrite $\phi _{k}\left( y\right)$ in a real form. The conclusion is that those fields can not be coupled via the $f(\chi)$ that appears in Eq. (\ref{fx100a1}).

In order to circumvent this issue, a generalization of Eq. (\ref{fx100a1}) is necessary. Regarding this, we have proposed an enlarged expression in terms of a new parameter $\beta$. The fields now decouple in a limit that involves $\beta$ and $\lambda$ simultaneously. This research may provide important insights about how different fields interact.

Preliminary simulations have indicated that, for $\lambda$ fixed and different $\beta$, the kink-antikink collision provides new interesting results. Nontrivial resonant structures are generated, with novel fractal patterns to be explored. We are currently organizing these results, and intend to present them in a future contribution.


\section*{Acknowledgements}

L. P. thanks the full support from Coordenação de Aperfeiçoamento de Pessoal de Nível Superior - CAPES (Brazilian agency, via a PhD scholarship) - Finance Code 001. E. H. acknowledges the partial financial support received from the Conselho Nacional de Desenvolvimento Científico e Tecnológico - CNPq (Brazilian agency) via the Grant N° 309604/2020-6.




\begin{thebibliography}{99}


\bibitem{prasom} M. K. Prasad and Charles M. Sommerfield, {\it Exact Classical Solution for the 't Hooft Monopole and the Julia-Zee Dyon}, {{Phy. Rev. Lett.} {\bf 35}, 12 (1975)}.

\bibitem{bogo} E. B. Bogomolny, {\it Stability of Classical Solutions}, {{Sov. J. Nucl. Phys.} {\bf 24}, 449 (1976)}.

\bibitem{masu} N. Manton and P. Sutcliffe, {\it{Topological Solitons}}, Cambridge University Press, (2004).

\bibitem{poga} S. D. Pollard et al., {\it{Bloch chirality induced by an interlayer Dzyaloshinskii-Moriya interaction in ferromagnetic multilayers}}, Phys. Rev. Lett. {\bf{125}}, 227203 (2020).

\bibitem{chdo} J. Chen and S. Dong, {\it{Manipulation of magnetic domain walls by ferroelectric switching: dynamic magnetoelectricity at the nanoscale}}, Phys. Rev. Lett. {\bf{126}}, 117603 (2021).

\bibitem{jab} P.-O. Jubert, R. Allenspach and A. Bischof, {\it{Magnetic domain walls in constrained geometries}}, Phys. Rev. B {\bf{69}}, 220410(R) (2004).

\bibitem{sugi} T. Sugiyama, {\it{Kink-antikink collisions in the two-dimensional $\phi^4$ model}}, Prog. Theor. Phys. {\bf{61}}, 1550 (1979).

\bibitem{camp1} D. K. Campbell, J. F. Schonfeld and C. A. Wingate, {\it{Resonance structure in kink-antikink interactions in $\phi^4$ theory}}, Physica D {\bf{9}}, 1 (1983).

\bibitem{anninos0} P. Anninos, S. Oliveira and R. A. Matzner, {\it{Fractal structure in the scalar $\lambda(\phi^2-1)^2 $ theory}}, Phys. Rev. D {\bf{44}}, 1147 (1991).

\bibitem{cape} D. K. Campbell and M. Peyrard, {\it{Solitary wave collisions revisited}}, Physica D {\bf{18}}, 47 (1986).

\bibitem{dohamerosh} P. Dorey et al., {\it{Boundary
scattering in the $\phi^4$ model}}, J. High Energy Phys. {\bf{05}}, 107 (2017).

\bibitem{asmoraebsa} A. Askari et al., {\it{Collision of $\phi^4$ kinks free of the Peierls-Nabarro barrier in the regime of strong discreteness}}, Chaos, Solitons and Fractals {\bf{138}}, 109854 (2020).

\bibitem{dorosh} P. Dorey, K. Mersh, T. Romanczukiewicz and Ya. Shnir, {\it{Kink-antikink collisions in the
$\phi^6$ model}}, Phys. Rev. Lett. {\bf{107}}, 091602 (2011).

\bibitem{weig} H. Weigel, {\it{Emerging translational variance: vacuum polarization energy of the
$\phi^6$ kink}}, Adv. High Energy Phys. {\bf{2017}}, 1486912 (2017).

\bibitem{mogasadmja} A. Moradi Marjaneh et al., {\it{Multi-kink collisions in the $\phi^6$ model}}, J. High Energy Phys. {\bf{07}}, 028 (2017).

\bibitem{blms} D. Bazeia, L. Losano, J. M. C. Malbouisson and J. R. L. Santos, {\it{Multi-sine-Gordon models}}, Eur. Phys. J. C {\bf{71}}, 1767 (2011).

\bibitem{masd} Aliakbar Moradi Marjaneh, Alidad Askari, Danial Saadatmand and Sergey V. Dmitriev, {\it{Extreme values of elastic strain and energy in sine-Gordon multi-kink collisions}}, Eur. Phys. J. B {\bf{91}}, 22 (2018).

\bibitem{pmrg} Marzieh Peyravi, Afshin Montakhab, Nematollah Riazi and Abdorrasoul Gharaati, {\it{Interaction properties of the periodic and step-like solutions of the double-Sine-Gordon equation}}, Eur. Phys. J. B {\bf{72}}, 269 (2009).

\bibitem{bgmsa} E. Belendryasova et al., {\it{A new
look at the double sine-Gordon kink-antikink scattering}}, J. Phys. Conf. Ser. {\bf{1205}}, 012007 (2019).

\bibitem {takyi} I. Takyi et al., {\it Kink collision in the noncanonical $\phi^6$ model: a model with localized inner structures}, {{Results in Physics} {\bf 44}, 106197 (2023)}.

\bibitem {henoli} T. Mendonça and H. de Oliveira, {\it The collision of two-kinks defects}, {{J. High Energy Phys.} {\bf 2015}, 120 (2015)}.

\bibitem {weigel} H. Weigel, {\it Kink-antikink scattering in $\phi^4$ and $\phi^6$ models}, {{J. Phys. Conf. Ser.} {\bf 482}, 012045 (2014)}.

\bibitem {gani1} V. A. Gani et al., {\it Scattering of the double sine-Gordon kinks}, {{Eur. Phys. J. C} {\bf 78}, 345 (2018)}.

\bibitem {dio1} Dionisio Bazeia, Jo\~ao G. F. Campos and Azadeh Mohammadi, {\it Kink-antikink collisions
in the $\phi^8$ model: short-range to long-range journey}, {{J. High Energy Phys.} {\bf 05}, 116 (2023)}.

\bibitem {simas1} F. C. Simas, Adalto R. Gomes, K. Z. Nobrega and J. C. R. E. Oliveira, {\it Suppression of two-bounce windows in kink-antikink collisions}, {J. High Energy Phys.} {\bf 2016}, 104 (2016).

\bibitem {dio2} D. Bazeia, E. Belendryasova and V. A. Gani, {\it Scattering of kinks of the sinh-deformed $\phi^4$ model}, {{Eur. Phys. J. C} {\bf 78}, 340 (2018)}.

\bibitem {dio3} A. Moradi Marjaneh, F. C. Simas and D. Bazeia, {\it Collisions of kinks in deformed $\phi^4$ and $\phi^6$ models}, {{Chaos, Solitons and Fractals} {\bf 164}, 112723 (2022)}.

\bibitem {halava.2012} A. Halavanau, T. Romanczukiewicz and Y. Shnir, {\it Resonance structures in coupled two-component $\phi^4$ model}, {{Phys.\ Rev.\ D} {\bf 86}, 085027 (2012)}.

\bibitem{Lima.JHEP.2019} F. C. Lima, F. C. Simas, K. Z. Nobrega and A. R. Gomes, {\it Boundary scattering in the $\phi^6$ model}, {{J. High Energy Phys.} {\bf 10}, 147 (2019)}.

\bibitem {ivan} Ivan C. Christov et al., {\it Kink-kink and kink-antikink interactions with long-range tails}, {{Phys. Rev. Lett.} {\bf 122}, 171601 (2019)}.

\bibitem{alonso} A. Alonso-Izquierdo, {\it{Non-topological kink scattering in a two-component scalar field theory model}}, Commun. Nonlin. Sci. Numer. Simul. {\bf{85}}, 105251 (2020).

\bibitem{alonso2} A. Alonso-Izquierdo, {\it{Reflection, transmutation, annihilation, and resonance in two-component kink collisions}}, Phys. Rev. D {\bf{97}}, 045016 (2018).

\bibitem{alonso3} A. Alonso-Izquierdo, {\it{Kink dynamics in the MSTB model}}, Phys. Scr. {\bf{94}}, 085302 (2019).

\bibitem{algomato} A. Alonso-Izquierdo, M. A. Gonz\'alez Le\'on, J. Mart\'in Vaquero and M. de la
Torre Mayado, {\it{Kink scattering in a generalized Wess-Zumino model}}, Commun. Nonlin. Sci. Numer. Simul. {\bf{103}}, 106011 (2021).

\bibitem{Halavanau} A. Halavanau, T. Romanczukiewicz and Ya. Shnir, {\it{Resonance structures in coupled two-component $\phi^4$ model}}, Phys. Rev. D {\bf{86}}, 085027 (2012).

\bibitem{simas2022} F. C. Simas, K. Z. Nobrega, D. Bazeia and A. R. Gomes, {\it{Asymmetry engendered by symmetric kink-antikink
scattering in a degenerate two-field model}},
Int. J. Mod. Phys. A {\bf{38}}, 2350083 (2023).

\bibitem{maevvaza} M. Mukhopadhyay, E. I. Sfakianakis, T. Vachaspati and G. Zahariade, {\it{Kink-antikink scattering in a quantum vacuum}}, J. High Energy Phys. {\bf{04}}, 118 (2022).

\bibitem{mavacha} M. Mukhopadhyay and T. Vachaspati, {\it{Resonance structures in kink-antikink scattering in a quantum vacuum}}, Phys. Rev. D {\bf{107}}, 116017 (2023).

\bibitem {adam.2019} C. Adam, K. Oles, T. Romanczukiewicz and A. Wereszczynski, {\it Spectral walls in soliton collisions}, {{Phys.\ Rev.\ Lett.} {\bf 122}, 241601 (2019)}.

\bibitem{camoquweza} J. G. F. Campos et al., {\it{Fermionic spectral walls in kink collisions}}, J. High Energy Phys. {\bf{01}}, 71 (2023).

\bibitem{adrowe1} C. Adam, K. Oles, T. Romanczukiewicz and A. Wereszczynski, {\it{Spectral walls in antikink-kink
scattering in the $\phi^6$ model}}, Phys. Rev. D {\bf{106}}, 105027 (2022).

\bibitem {adam.2021} C. Adam et al., {\it Spectral walls in multifield kink dynamics}, {{J. High Energy Phys.} {\bf 08}, 147 (2021)}.

\bibitem{blm} D. Bazeia, M. A. Liao and M. A. Marques, {\it{Geometrically constrained kinklike configurations}}, Eur. Phys. J. Plus {\bf{135}}, 383 (2020).

\bibitem{balbazmar} A. J. Balseyro, D. Bazeia and M. A. Marques, {\it{Mechanism to induce geometric constriction on kinks and domain walls}}, Eur. Phys. Lett. {\bf{141}}, 34003 (2023).

\bibitem{bazmarmen} D. Bazeia, M. A. Marques and R. Menezes, {\it{Geometrically constrained kink-like configurations engendering long-range, double-exponential, half-compact and compact behavior}}, Eur. Phys. J. Plus {\bf{138}}, 735 (2023).

\bibitem{marmen} M. A. Marques and R. Menezes, {\it{Geometrically constrained multifield models with BNRT solutions}}, Chaos, Solitons and Fractals {\bf{181}}, 114730 (2024).

\bibitem{hiatma} M. C. Hickey, D. Atkinson, C. H. Marrows and B. J. Hickey, {\it{Controlled domain wall nucleation and resulting magnetoresistance in $Ni_{81}Fe_{19}$ nanoconstrictions}}, J. of Applied Phys. {\bf{103}}, 07D518 (2008).

\bibitem{clahu} D. Claudio-Gonzalez et al., {\it{Fabrication and simulation of nanostructures for domain wall magnetoresistance studies on nickel}}, J. Magn. Magn. Mater. {\bf{322}}, 1467 (2010).

\bibitem{chego} A. P. Chen, J. Gonzalez and K. Y. Guslienko, {\it{Domain walls confined in magnetic nanotubes with uniaxial anisotropy}}, J. Magn. Magn. Mater. {\bf{324}}, 3912 (2012).

\bibitem{thivan} J. M. Thijssen and H. S. J. Van der Zant, {\it{Charge transport and single-electron effects in nanoscale systems}}, Phys. Status Solidi B {\bf{245}}, 1455 (2008).

\bibitem{fermion} D. Bazeia, A. Mohammadi and D. C. Moreira, {\it{Fermions in the presence of topological structures under geometric constrictions}}, Phys. Rev. D {\bf{103}}, 025003 (2021).

\bibitem{joao} João G. F. Campos, Fabiano C. Simas and D. Bazeia, {\it{Kink scattering in the presence of geometric constrictions}}, J. High Energy Phys. {\bf{10}}, 124 (2023).

\bibitem{barisa} D. Bazeia and M. M. Santos, {\it{Classical stability of solitons in systems of coupled scalar fields}}, Phys. Lett. A {\bf{217}}, 28 (1996).

\bibitem{banari} D. Bazeia,
H. Boschi-Filho and F. A. Brito, {\it{Domain defects in systems of two real scalar fields}}, J. High Energy Phys. {\bf{4}},
028 (1999).

\end{thebibliography}
\end{document}